\documentclass[final,5p,times, twocolumn]{elsarticle}
\usepackage[T1]{fontenc}
\usepackage{microtype,graphicx}
\graphicspath{{figures/}}
\biboptions{sort&compress}
\usepackage{amsmath,amssymb,bm,booktabs,tabularx,array}
\usepackage{placeins}
\usepackage[hidelinks]{hyperref}

\newcommand{\kk}{\mathbf{k}}
\newcommand{\xx}{\mathbf{x}}
\newcommand{\vv}{\mathbf{v}}
\newcommand{\EE}{\mathbf{E}}
\newcommand{\vth}{v_{\mathrm{th}}}
\newcommand{\vb}{v_b}
\newcommand{\CIC}{\mathrm{CIC}}
\newcommand{\Rg}{R_\gamma}
\newcommand{\sinc}{\operatorname{sinc}}

\newcolumntype{L}{>{\raggedright\arraybackslash}X}
\begin{document}
\hypersetup{pdfauthor={Ayushi Agrawal, Shan-Chang Lin, Ezhilmathi Krishnasamy, Pascal Bouvry}}
\begin{frontmatter}
\title{Dynamics-aware bandwidth selection for smoothed charge deposition in 2D electrostatic particle-in-cell}
\author[lux]{Ayushi Agrawal\corref{corresponding}}
\ead{ayushi.agrawal@uni.lu}
\author[lux]{Shan-Chang Lin}
\author[lux,lj,rud]{Ezhilmathi Krishnasamy}
\author[lux]{Pascal Bouvry}
\cortext[corresponding]{Corresponding author.}
\address[lux]{Faculty of Science, Technology and Medicine, University of Luxembourg, Esch-sur-Alzette, Luxembourg}
\address[lj]{Faculty of Mechanical Engineering, University of Ljubljana, Ljubljana, Slovenia}
\address[rud]{Rudolfovo-Science and Technology Centre Novo Mesto, Slovenia}

\begin{abstract}
Charge deposition in the particle-in-cell (PIC) method can be viewed as kernel density estimation on a spatial mesh. This allows the smoothing bandwidth to be selected through statistical criteria such as mean integrated squared error or cross-validation, but smoothing also changes the electric field acting on the particles. Therefore, a selected bandwidth can alter the growth or damping of physical modes. Here, we develop a dynamics-aware bandwidth-selection procedure that combines a density-error objective with constraints on specified linear modes. These constraints follow from the plasma equilibrium and the combined transfer, and limit the predicted change in the selected rates. In this work, we demonstrate the need for equilibrium information through a controlled two-stream experiment that varies the beam drift while holding the initial spatial density, thermal samples, wavevector and fixed-filter transfer unchanged. With this filter held fixed, the measured growth-rate reduction relative to CIC increases from $0.72\%$ to $9.11\%$ across four resolved equilibria as marginal stability is approached. We then apply the constrained selection rule to 115 Gaussian and compensated-binomial candidates. Density error alone selects a raw binomial filter that removes the protected growth, in the model and in a projected-linear test. The constraint applied here, selects a degree four compensated binomial filter, which lowers the held-out density and grid-field mean-square errors by 76.4\% and 37.6\%. While, In a separate projected-linear test, the ratio of the filtered growth rate to the CIC growth rate for the same mode is  0.9595, with a paired 95\% confidence interval of [0.9500, 0.9670]. This interval contains the model prediction and lies within the 10\% tolerance. Projected-linear Landau calculations reproduce the predicted damping ratios to within 0.2\% of a resolved baseline. We also analyse finite-marker loading and ensemble averaging to relate the model response to the measured rates. Together, these results demonstrate a filter-design procedure that combines statistical error reduction with subsequent validation of the selected linear dynamics. The proposed procedure is useful for noise reduction for measurements of weak instabilities near marginal stability and studies of linear wave damping, where modest changes in particle-field coupling can affect the physical rates. 

\end{abstract}


\end{frontmatter}

\section{Introduction}
\label{Section1}

The kinetic evolution of a collisionless plasma is described by the distribution function $f_s(\xx,\vv,t)$ of each species. It evolves according to the Vlasov equation, coupled self-consistently to the electromagnetic fields. A direct Eulerian discretisation uses a six-dimensional phase-space mesh in the general three-dimensional case. If $N_{\mathrm{grid}}$ points are used in each direction, the number of phase-space cells scales as $N_{\mathrm{grid}}^6$~\cite{birdsall2004plasma,hockney1988computer}. 

The particle-in-cell (PIC) method reduces this cost by representing $f_s$ with a finite ensemble of computational macro-particles. The particles follow the characteristics of the Vlasov equation, while the self-consistent fields are computed on a spatial mesh~\cite{dawson1983review,birdsall2004plasma,hockney1988computer}. Particle operations scale linearly with the number of macro-particles $N_p$, while the field-solver cost depends on the chosen discretisation; for example, an fast-Fourier-transform (FFT) Poisson solve scales as $\mathcal{O}(N_c\log N_c)$ for $N_c$ mesh cells~\cite{hockney1988computer}. This favourable scaling and efficient parallelisation have made PIC a standard method for kinetic plasma simulation~\cite{dawson1962onedimensional,dawson1983review,verboncoeur2005particle}. PIC and its reduced kinetic variants are used for gyrokinetic turbulence in magnetic confinement~\cite{lee1983gyrokinetic,parker1993fully,lin1998turbulent,garbet2010gyrokinetic}, laser-plasma interactions and plasma-based acceleration~\cite{tajima1979laser,fonseca2002osiris,derouillat2018smilei,arber2015contemporary,vay2018warpx}, magnetic reconnection and relativistic collisionless shocks~\cite{daughton2006fully,spitkovsky2008particle}, and low-temperature discharges modelled with Monte Carlo collisions~\cite{vahedi1995monte,turner2013simulation}. In all of these applications, physical fidelity depends directly on how particles and mesh fields are coupled.

A fundamental limitation of PIC is statistical noise~\cite{langdon1979kinetic,nevins2005discrete,qin2018kde}. Since $f_s$ is represented by a finite sample, estimated moments such as density, current, temperature and heat flux generally contain sampling error. For independent loading at fixed mesh resolution, its amplitude typically decreases as $1/\sqrt{N_{\rm ppc}}$, where $N_{\rm ppc}$ is the number of particles per cell~\cite{langdon1979kinetic,hockney1988computer,holod2007statistical,nevins2005discrete}. Thus, reducing the sampling error by one order of magnitude requires approximately two orders of magnitude more particles. Noise in velocity moments also affects derived quantities and their spatial gradients~\cite{picklo2023siac}. In addition, aliasing, finite-grid instabilities, and grid heating can become important when the Debye length is poorly resolved~\cite{langdon1970effects,okuda1972nonphysical,huang2016finite,barnes2021finite,werner2025smoothing}. Both sampling noise and grid-related errors depend on the deposition and gathering operators. Common deposition functions include nearest-grid-point, cloud-in-cell (CIC), triangular-shaped-cloud and higher-order spline shapes~\cite{birdsall2004plasma,langdon1970effects,lewis1970energy,langdon1973energy}. 

Several approaches have been used to reduce this statistical cost of PIC. The $\delta f$ method evolves a perturbation around a known background to reduce the associated fluctuations with the full distribution representation~\cite{parker1993fully,hu1994generalized}. The Ensemble averaging method combines independent realisations to improve the resolution of a common signal~\cite{touati2022ensemble}. Higher-order shapes and digital filters have been used to suppress selected spatial scales, while sparse-grid methods provide an alternative way to balance particle and mesh errors~\cite{birdsall2004plasma,muralikrishnan2020sparse}. These approaches act on different parts of the calculation. Here we focus on the bandwidth selection applied within the deposition--field--gather coupling. A broader, smoother kernel suppresses short-wavelength fluctuations and aliasing, but it also attenuates physical gradients and modes~\cite{okuda1972nonphysical,werner2025smoothing}. The bandwidth therefore controls a bias--variance trade off. This trade off is commonly measured by the mean integrated squared error (MISE)~\cite{rosenblatt1956remarks,parzen1962estimation,silverman1986density,wand1995kernel}. A bandwidth can then be chosen by minimising an estimated error, using a plug-in rule or cross-validation when the underlying density is unknown. This gives a systematic statistical objective for particle shapes and mesh filters. Such selection procedures have already produced useful advances in PIC. Wu and Qin introduced a von Mises deposition shape with widths selected through cross-validated MISE, obtaining improved density estimates and Landau-damping measurements in their benchmarks~\cite{qin2018kde}. Evstatiev et al. developed noise and error analyses for particle-based plasma calculations, including bias--variance optimisation and charge-conserving shapes of arbitrary width~\cite{evstatiev2021noise}. These studies establish the value of treating interpolation width as a numerical parameter that can be selected from the data. 

The effect of particle shapes on discrete dispersion and finite-grid instability is likewise well established~\cite{langdon1970effects,okuda1972nonphysical,huang2016finite,barnes2021finite}. Werner et al. showed how smoothing integrated with the field solve can suppress noise and grid instability in electrostatic PIC~\cite{werner2025smoothing}. For a calculation designed to follow a specified physical mode, an additional objective is to retain its growth or damping rate. Charge conservation and reduced density fluctuations are valuable properties, but neither determines the rate. The relevant response also depends on the equilibrium velocity distribution and the numerical particle to field operator. This dependence is transparent for a matched, zero-phase smoother. When the same filter is applied to the deposited density and to the field before gathering, its Fourier transfer enters the coupling twice~\cite{langdon1970effects,huang2016finite,werner2025smoothing}. The resulting round-trip factor modifies the dielectric response and shifts the roots of the dispersion relation, as derived in Section~\ref{Section2}. For the branches examined here, reduced coupling lowers two-stream growth and increases the magnitude of Landau damping. The relative change can be especially significant near marginal stability, where the unsmoothed growth rate is small. A small change in coupling can then represent a substantial fraction of the mode's rate.

Moreover, two equilibria may have identical spatial densities and different velocity distributions. An objective evaluated only from their initial density samples then receives the same information. However, the growth or damping of a plasma wave depends on the velocity distribution as well as the density. The same filter can consequently produce different changes in the wave’s rate in the two equilibria. Therefore, in the present work, we incorporate this missing information in two steps. First, determine which bandwidths preserve the modes of interest by using the equilibrium velocity distribution and the implemented particle-field coupling to predict how smoothing changes their growth or damping rates. Then minimise the density error among those admissible bandwidths. Its central aim is to retain the statistical error objective while retaining the equilibrium information that tells how the filter changes the plasma response.

In this work, we test the response constraint in two-dimensional, two-velocity electrostatic calculations. The central experiment varies the beam drift in a two-stream equilibrium at a fixed oblique wavevector. Particle positions, thermal samples, perturbation and the fixed control filter remain unchanged across the experiment. This design isolates the equilibrium dependence of the response from changes in initial density or filter attenuation. Across four resolved equilibria, the fixed Gaussian's reduction of the mean individual-seed growth rate increases from approximately $0.72\%$ to $9.11\%$. The selected interior settings instead retain rate ratios between $0.9504$ and $0.9545$, with paired $95\%$ confidence intervals inside the prescribed $10\%$ tolerance. The accepted linear measurements use a mode-projected $\delta f$ formulation described in Section~\ref{Section3}, built on perturbation-weight methods~\cite{parker1993fully,hu1994generalized}.

We then apply the complete constrained search jointly to 115 candidates, CIC, 84 Gaussians and 30 compensated binomials, evaluated offline on common particle data with separate selection and evaluation seeds. The constraint selects a degree four compensated binomial filter, which reduces the density and grid-field mean-square errors by 76.4\% and 37.6\%, against 7.2\% and 3.5\% for the best admissible Gaussian. In a separate projected-linear calculation its growth-rate ratio to CIC is 0.9595, consistent with the prediction within the paired confidence interval. These reconstruction errors are measured on snapshots evolved with CIC, whereas rate preservation is assessed in a separate projected-linear calculation.
Together, these studies examine the physical information required for selection and the error reduction retained by the chosen setting. The central contribution of this work is a dynamics-aware constraint on statistical bandwidth selection. The controlled drift scan establishes the need for that constraint, and the exact-width experiment demonstrates statistical selection followed by response validation. The finite-marker and ensemble analyses explain which measured rates can be compared with the response model. It connects the statistical objective to the implemented coupling and the modes of interest.

Section~\ref{Section2} introduces the matched filter, derives its effect on the linear response, and formulates the selection procedure. Section~\ref{Section3} describes the numerical experiments and rate measurements. Section~\ref{section4} presents the results, followed by the conclusions in Section~\ref{Section5}. In the appendices, we present the susceptibility and conservation checks. Particle loading, discrete-response derivations and detailed experimental results are provided in the supplementary material.

\section{Theory and selection method}
\label{Section2}
\subsection{Filtered particle--field coupling}
We consider the collisionless Vlasov--Poisson equations~\cite{birdsall2004plasma,hockney1988computer},
\begin{equation}
\begin{gathered}
 \partial_t f_s+\vv\cdot\nabla_\xx f_s+
 \frac{q_s}{m_s}\EE\cdot\nabla_\vv f_s=0,\\
 -\nabla^2\phi=\rho/\varepsilon_0,\qquad \EE=-\nabla\phi.
\end{gathered}
 \label{Equation1}
\end{equation}
where $\xx$, $\vv$ and $t$ denote position, velocity and time, $f_s$ is the phase-space number density of species $s$ with charge $q_s$ and mass $m_s$, and $\phi$ is the electrostatic potential. We use two spatial and two velocity dimensions (2D2V), with periodic boundaries and a neutralising fixed ion background. The charge density $\rho=\sum_s q_s\int f_s\,d^2v+\rho_{\rm ion}$, is deposited with CIC weights. For domain lengths $L_i$ and grid counts $N_i$, the wavevector is $k_i=2\pi m_i/L_i$, where $m_i$ is an integer mode index and $\Delta_i=L_i/N_i$ ($i=x,y$). Fourier transforming Poisson's equation and $\EE=-\nabla\phi$ gives, for $\kk\ne0$,
\begin{equation}
 \widehat{\EE}(\kk)=-\frac{i\kk}{\varepsilon_0|\kk|^2}\widehat\rho(\kk).
 \label{Equation2}
\end{equation}

We smooth the mesh quantities with a Gaussian whose widths $\sigma_x$ and $\sigma_y$ are fixed in space and time. Its covariance is $H=\operatorname{diag}(\sigma_x^2,\sigma_y^2)$, and the corresponding convolution operator $\mathsf S_H$ is real and even, so it introduces no phase shift. The Fourier transform of the normalised Gaussian gives~\cite{silverman1986density,wand1995kernel}
\begin{equation}
 W_H(\kk)=\exp[-\kk^{\mathsf T}H\kk/2].
 \label{Equation3}
\end{equation}
The same Gaussian is applied to the deposited density before the field solve, and one before it is gathered to the particles. To write this matched operation, let $C_{gp}$ be the dimensionless CIC weight between particle $p$ and node $g$, and let $\Delta V=\Delta_x\Delta_y$. Without additional smoothing, deposition and gathering are
\begin{equation}
 \rho_g=\frac{1}{\Delta V}\sum_p C_{gp}Q_p,
 \qquad \EE_p=\sum_g C_{gp}\EE_g.
 \label{Equation4}
\end{equation}
In matrix form, deposition is $\mathsf B=C/\Delta V$ and gathering is $\mathsf B^\dagger=\Delta V\mathsf B^{\mathsf T}=C^{\mathsf T}$, the adjoint under the mesh-volume inner product. Including the two filtering steps gives
\begin{equation}
 \EE_p=\mathsf B^\dagger\mathsf S_H\mathsf A\mathsf S_H\mathsf B\,\mathbf Q,
 \label{Equation5}
\end{equation}
where $\mathsf A$ is the spectral charge-to-field map and $\mathbf Q=(Q_p)$ is the vector of marker charges. In Fourier space, the diagnostic field between the solve and return filtering contains one factor $W_H$ and the self-consistent particle coupling contains $W_H^2$.

The Fourier transform of the CIC shape is the square of a sinc function~\cite{hockney1988computer,langdon1970effects}. The principal spatial harmonic therefore has transfer
\begin{equation}
 S_\CIC(\kk)=\prod_{i=x,y}\sinc^2(k_i\Delta_i/2).
 \label{Equation6}
\end{equation}
Here $\sinc z=\sin z/z$ with $\sinc 0=1$. For mode indices $m_i$, grid sizes $N_i$ and cell-normalised widths $h_i=\sigma_i/\Delta_i$, the round-trip principal-harmonic factor is
\begin{equation}
\begin{gathered}
 \mathcal T_H(\mathbf m)=S_\CIC^2W_H^2
 =e^{-q}\prod_{i=x,y}\sinc^4(\pi m_i/N_i),\\
 q=\sum_{i=x,y}(2\pi m_i h_i/N_i)^2.
\end{gathered}
 \label{Equation7}
\end{equation}
The logarithmic attenuation is $q=-\ln W_H^2=\kk^{\mathsf T}H\kk$. Setting $q=0$ recovers CIC without additional smoothing. A bound $q\le q_{\max}$ therefore limits attenuation of the selected mode.

The matched construction preserves represented charge because CIC weights form a partition of unity and $W_H(0)=1$. With $d=\mathsf S_H\mathsf B\mathbf Q$, the internal-force sum is proportional to $d^{\mathsf T}\mathsf A d=0$. The corresponding single-particle quadratic form gives zero self-force~\cite{brackbill2016conservation}. 

\subsection{Equilibrium sensitivity of a filtered mode}
For perturbations proportional to $\exp[i(\kk\cdot\xx-\omega t)]$, let $\omega=\omega_r+i\gamma$, where $\omega_r$ is the oscillation frequency and $\gamma$ is the growth/ damping rate. Linearising Eq.~\eqref{Equation1}, integrating over velocity and substituting the perturbed charge into Poisson's equation gives the principal-harmonic dispersion relation~\cite{langdon1970effects,huang2016finite,werner2025smoothing},
\begin{equation}
 D_H(\kk,\omega)=1+\mathcal T_H(\kk)\chi(\kk,\omega)=0,
 \label{Equation8}
\end{equation}
where $\chi$ is the susceptibility of the equilibrium velocity distribution. The factor $\mathcal T_H$ is the numerical coupling, while $\chi$ describes how the plasma responds to the field. Changing the filter changes the roots of their combined dispersion relation. Equation~\eqref{Equation8} is the principal-harmonic relation that explains the mechanism, but a finite mesh and time step can also introduce spatial and temporal aliases~\cite{langdon1970effects,langdon1979time}. We include these relations in the numerical response to set the two-stream filter widths. ~\ref{A1} and Supplementary Section~S5 derives the response for the implemented numerical response in detail.

For a simple root of $D(\omega,a)=0$, with $a=W_H^2$ and $\partial_\omega D\ne0$, differentiation gives
\begin{equation}
 \frac{d\omega}{da}=-\frac{\partial_aD}{\partial_\omega D}.
 \label{Equation9}
\end{equation}
Therefore, the response to a change in attenuation depends on the equilibrium through both derivatives. Moreover, a given absolute shift has a larger relative effect when $|\gamma|$ is small. Thus, a useful limiting case is a marginal branch with a nondegenerate double root at $\omega=0$ and coupling $a_c$: $D=\partial_\omega D=0$, with nonzero $\partial_aD$ and $\partial_\omega^2D$. 

The Taylor expansion of $D$ in neighbourhood of $a_c$ gives $(\partial_aD)(a-a_c)+\tfrac12(\partial_\omega^2D)\omega^2$. Setting $\omega=i\gamma$ and differentiating this relation gives
\begin{equation}
 \frac{\Delta\gamma}{\gamma}\simeq\frac{\Delta a}{2(a-a_c)}.
 \label{Equation10}
\end{equation}
The denominator explains the sensitivity. If the equilibrium is far from its marginal coupling \(a_c\), a small change in \(a\) has a modest relative effect. If \(a\) is close to \(a_c\), the same change can produce a much larger percentage change in $\gamma$. Equation~\ref{Equation10} is a local result. It requires the stated double-root geometry and \(|\Delta a|\ll|a-a_c|\). In the actual selection calculation, we use branch continuation to determine the actual tolerance boundary. This explains why one attenuation level can produce different relative rate changes in different equilibria. For the branches considered here, the computed roots show decreasing two-stream growth and increasing weak-Landau damping magnitude as the coupling is reduced.

\subsection{Equilibrium-aware filter selection}

The purpose of the constraint is to retain the linear modes relevant to a given calculation. We call these protected modes and denote their wavevectors by $\mathcal{M}_{\mathrm{phys}}$.
Let $\Theta$ be a finite set of candidate filters, including unsmoothed CIC. Each candidate $\theta$ specifies the filter family and its parameters: the covariance $H$ for a Gaussian, or the compensation degree $p$ and pass count $n$ for a compensated binomial filter. Its one-way transfer is $W_\theta(\mathbf{k})$; for the real, symmetric filters considered here, matched filtering gives the coupling factor $W_\theta^2(\mathbf{k})$.

For each candidate, we compare the filtered rate with the CIC rate. The rate ratio and admissible set are
\begin{equation}
\begin{aligned}
R_\gamma(\mathbf{k};\theta) = \frac{\gamma_\theta(\mathbf{k})}{\gamma_0(\mathbf{k})},\\
\mathcal{A}_\epsilon = \bigl\{\theta\in\Theta:
 |R_\gamma(\mathbf{k};\theta)-1|\leq\epsilon\hspace{1.3em}\text{for all }
 \mathbf{k}\in\mathcal{M}_{\mathrm{phys}}\bigr\}.
\end{aligned}
\label{Equation11}
\end{equation}
Here, $\gamma_0$ and $\gamma_\theta$ are obtained from the same response model by continuing the CIC branch as the filter transfer changes. The baseline rate must be nonzero, and the branch must remain identifiable. Candidates with unresolved roots are excluded from selection. For $0<\epsilon<1$, the constraint excludes a zero rate or a reversal of its sign.

Among the admissible candidates, we select a filter that minimises the estimated density error at the operating tolerance $\epsilon_{\mathrm{op}}$:
\begin{equation}
\theta_*\in
\mathop{\mathrm{arg\,min}}_{\theta\in\mathcal{A}_{\epsilon_{\mathrm{op}}}}
\widehat{\mathcal{E}}_\rho(\theta).
\label{Equation12}
\end{equation}
The density-error objective is
\begin{equation}
\begin{aligned}
\widehat{\mathcal{E}}_\rho(\theta)
&=\frac{1}{N_rN_tN_g}\sum_{r,j,g}\bigl[\rho_{\theta,r}(\mathbf{x}_g,t_j)
-\rho_{\mathrm{ref},r}(\mathbf{x}_g,t_j)\bigr]^2.
\end{aligned}
\label{Equation13}
\end{equation}
The average covers $N_r$ particle realisations, $N_t$ saved
times and $N_g$ mesh nodes. The reference is a paired calculation
with higher particle count. Selection and evaluation use
disjoint realisations.

Writing $\theta_0$ for unsmoothed CIC, the density-error
reduction factor is
$\mathcal{N}_\rho=
\widehat{\mathcal{E}}_\rho(\theta_0)/
\widehat{\mathcal{E}}_\rho(\theta)$.
The corresponding field-error factor $\mathcal{N}_E$ uses the
squared Euclidean field difference, with the nodal field
evaluated before return filtering and gather. The fractional
error reduction is $1-1/\mathcal{N}$. These quantities compare
mean squared errors against a finite-particle reference;
they are not pure variance-reduction factors.
Cross-validation could be used as an alternative objective,
but its constrained use with the present particle loading
would require a separate assessment.

\subsection{Model boundary and selection procedure}
For the growing two-stream mode, we use $\varepsilon_{\rm acc}=0.10$ as the model's tolerance boundary. It follows the same dispersion root as in the CIC case. Similarly, we select an interior target as $\Rg=0.95$, corresponding to $\varepsilon_{\rm op}=0.05$ on the decreasing branch. This margin is a specified operating choice. For constrained error minimisation, the operational procedure is:
\begin{enumerate}
 \item Specify the equilibrium, implemented field coupling, protected modes, candidate set, error metric and both tolerances.
 \item Compute the CIC roots and continue the same branches as the exact filter transfer changes.
 \item Determine each mode's admissible attenuation at $\varepsilon_{\rm op}$ and form the intersection over the protected set.
 \item Evaluate the specified error objective on the feasible candidates and select its minimum, using selection data separate from the evaluation data.
 \item Evaluate the selected setting on independent seeds using paired uncertainty, signal-duration checks and the acceptance tolerance $\varepsilon_{\rm acc}$.
\end{enumerate}
In the drift scan, we test the response targeting by solving $\Rg=0.95$ along $h_x=0$, without re-minimising the density error at each drift. We also apply Eq.~\eqref{Equation12} to the 115-candidate experiment with $\epsilon_{\mathrm{op}}=0.05$. The selected candidate is evaluated on disjoint statistical samples and then tested with the selected filter active in projected-linear dynamics (Section~3).

To illustrate how the allowed attenuation depends on the equilibrium, we consider an oblique two-stream benchmark with two equally weighted Maxwellian beams drifting at $\pm v_b\mathbf e_x$, where $\mathbf e_x$ is the unit vector along the beam direction. We examine the mode $(m_x,m_y)=(12,1)$ on a $128\times16$ mesh in a domain of size $L_x\times L_y=64\times16$. The thermal standard deviation is $v_{\rm th}=0.05$. At $v_b=0.80$, substituting $m_i/N_i$ into Eq.~\eqref{Equation7} gives the numerical 10\% boundary, with its attenuation limit rounded to three significant figures,
\begin{equation}
 q=0.34698h_x^2+0.15421h_y^2\lesssim0.00909.
 \label{Equation14}
\end{equation}

 \begin{figure*}[t]
\centering\includegraphics[width=\textwidth]{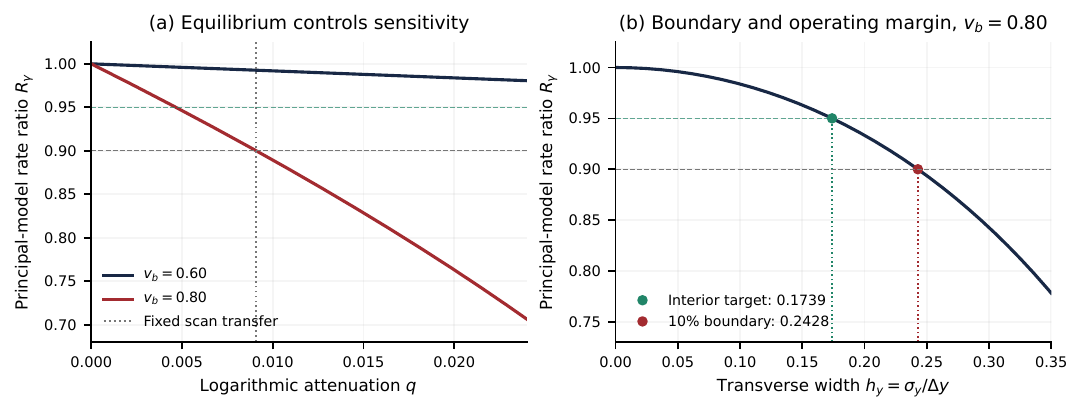}
\caption{Continuous-time principal-harmonic prediction for the oblique two-stream mode. (a) Growth-rate ratio versus Gaussian attenuation $q$ at two beam drifts. (b) At $v_b=0.80$, the ratio versus transverse width, with the 10\% tolerance boundary and 0.95 interior target.}
\label{Figure1}
\end{figure*}

 With $h_x=0$, its transverse width is $h_y=0.242735$, whereas the numerical 0.95-target setting is $h_y=0.173859$. The upper limit $0.00909$ is found by continuing the numerical dispersion root to $\Rg=0.90$. In Figure~\ref{Figure1} (a), we plot the predicted rate ratio against attenuation \(q\) at two beam drifts. The dotted vertical line marks the same fixed filter transfer in both cases. It intersects the \(v_b=0.80\) response near the 10\% boundary but has a much smaller predicted effect at \(v_b=0.60\). Changing the equilibrium changes the mode’s sensitivity even though the filter is fixed. In Figure~\ref{Figure1} (b), we show the \(v_b=0.80\) response with transverse width \(h_y\), assuming \(h_x=0\). Its two marked points show the 10\% boundary and the 5\% interior target. These curves are predictions from a principal-harmonic response model.

\subsection{Compensated binomial transfer}

The preceding construction uses a Gaussian width to control attenuation, but the constraint itself requires only the filter transfer. This allows us to examine filters with a different balance between preserving long wavelengths and smoothing short wavelengths. Binomial filters are a useful example as their strength is set by a pass count, while compensation modifies the attenuation near long-wavelength modes we wish to retain. We derive their transfer here so that pass count and compensation order can be assessed with the same dispersion relation. 

For a binomial stencil $(1,2,1)/4$, the one-pass Fourier response is
$
\frac{e^{-ik\Delta}+2+e^{ik\Delta}}{4}
=\cos^2(k\Delta/2)=1-s,\quad s=\sin^2(k\Delta/2),
$
where $k$ and $\Delta$ are the wavenumber and grid spacing in the filtered direction~\cite{birdsall2004plasma,hockney1988computer}. Applying the stencil $n$ times gives $B_n=(1-s)^n$. To reduce attenuation in the low-wavenumber passband, we multiply $B_n$ by the Taylor polynomial of its inverse, $(1-s)^{-n}$, through degree $p$. The binomial-series identity gives the resulting order-$p$ compensated response:
\begin{equation}
\begin{aligned}
 W_{p,n}(k)&=(1-s)^n\sum_{j=0}^{p}\binom{n+j-1}{j}s^j\\
 &=1+\mathcal O(s^{p+1}).
\end{aligned}
 \label{Equation15}
\end{equation}
The compensator cancels the terms proportional to $s,\ldots,s^p$ in the raw-filter error. Since $s=\mathcal O((k\Delta)^2)$ as $k\Delta\to0$, the leading deviation from unity is of order $(k\Delta)^{2p+2}$. Compensation therefore preserves long-wavelength modes more closely while retaining attenuation at shorter wavelengths. 
For separable filtering, the directional responses multiply and their product is squared in the matched coupling. Substituting this exact transfer for $W_H$ applies the same rate constraint to compensation order and pass count. At finite wavenumber, and for compensated filters, the response constraint is evaluated using the exact transfer in Equation~\eqref{Equation15}. The mode-dependent exact Gaussian equivalence is given in ~\ref{A1}.

\section{Numerical experiments}
\label{Section3}
\subsection{Benchmark configurations and particle loading}
The experiments distinguish the impact of smoothing on a chosen plasma mode from its impact on the accuracy of the deposited quantities. We use periodic 2D2V electrostatic calculations with a neutralising ion background and units $\omega_p=n_0=\varepsilon_0=m_e=|q_e|=1$. The two-stream equilibrium consists of equally weighted Maxwellians drifting at $\pm v_b\mathbf e_x$, each with thermal standard deviation $\vth=0.05$. The weak-Landau equilibrium is a Maxwellian with $\vth=1$. Table~\ref{Table1} summarises the configurations; full parameter lists and sample assignments are given in Supplementary Section~S2. 
\begin{table*}[t]
\centering\small
\caption{Numerical configurations and their measurement roles.}
\label{Table1}
\begin{tabularx}{\textwidth}{@{}lLLL@{}}\toprule
Configuration & Domain and mesh & Evolution & Measurement\\\midrule
Two-stream response & $(64,16)$; $128\times16$ & Projected linear $\delta f$; PPC 512; $\Delta t=0.02$ & Growth of $(12,1)$ with the candidate filter active\\\addlinespace
Two-stream reconstruction & $(64,16)$; $128\times16$ & Full $f$ with CIC; PPC 32; $\Delta t=0.05$ & Offline density and field error over $[0.5,4]$; reference PPC 1024\\\addlinespace
Weak Landau & $(5\pi,8)$; $64\times16$ & Full $f$; PPC 1024; $\Delta t=0.05$ & Decay of $(1,0)$, $k\lambda_D=0.4$; error reference PPC 4096\\\addlinespace
Landau response & $(5\pi,8)$; $64\times16$ & Projected linear $\delta f$; PPC 1024; $\Delta t=0.05$ & Damping of $(1,0)$ to $t=70$\\\bottomrule
\end{tabularx}
\end{table*}

To isolate two-stream growth, we write $f=f_0+\delta f$ and linearise the evolution about the prescribed equilibrium. Markers sample $f_0$ and move at fixed velocities, while signed weights represent the perturbation driven by the electric field~\cite{parker1993fully,hu1994generalized,denton1995bounded}. The weights are advanced by explicit midpoint. Before each field solve, we retain only the target Fourier mode and its conjugate, which together represent a real perturbation. Velocities are loaded from global inverse normal quantiles, with independent permutations in the two coordinates and balanced beam assignments for the two-stream equilibrium. The loading and update equations are detailed in Supplementary Sections~S3-S4.

The drift scan uses $v_b=0.60,0.70,0.76,0.80,0.818$. Within each seed, spatial positions, charges, thermal samples, beam assignments and the initial perturbation are kept constant. Thus, the initial density and its statistical selection objective stay fixed while the susceptibility changes. The fixed Gaussian has $(h_x,h_y)=(0,0.2427346)$ and $W_H^2=0.99095494$ at every drift. A second setting is calculated separately at each equilibrium to target $\Rg=0.95$.

\subsection{Filter selection and comparisons}
At $v_b=0.80$, we compare CIC with 84 Gaussian filters spanning four attenuation levels and 21 allocations of smoothing between the two directions. Five independent particle realisations, distinguished by their random seeds, are used to determine the width to evaluate error. These states evolve a $(12,0)$ perturbation with unfiltered CIC coupling. The CIC comparator is the unfiltered deposition of those states. This comparison checks how well each filter reconstructs the density and field from the same particle data. To measure the change in dynamics, we then run a separate projected-linear calculation with the selected filter active throughout the evolution of the protected $(12,1)$ mode.

The compensated-binomial family contains 30 filters, with compensation degree $p=0,\dots,4$ ($p=0$ is the raw binomial) and pass count $n=1,2,3,4,6,8$, using the same filter in both directions. Its density and field errors are evaluated on the same particle states and seeds as the Gaussian candidates, so the two families are compared on a common sample. Rates are measured for the selected binomial filter and for the unconstrained density optimum (the candidate with the smallest selection density error when the constraint is omitted) on the drift-scan seeds, and for $(p,n)=(1,1),(2,2),(2,4),(3,4)$ on a separate ten-seed ensemble, see Table~\ref{Table1}. All rate comparisons are paired. The four representative binomials are paired with the CIC runs of their own ensemble. The Landau rate comparisons are paired in the same way.
 
The two-stream calculation applies CIC, the selected Gaussian, the selected $(4,8)$ filter, with velocity Verlet and the matched operator of Eq.~\eqref{Equation5}. It seeds the protected $(12,1)$ mode with amplitude $10^{-3}$ and uses the velocity loading above at two particle counts. The Landau comparison uses 40 paired seeds with three settings: CIC, a one-cell Gaussian $(1,1)$ and a wider Gaussian $(7.2443,7.2443)$.
 
\subsection{Rate measurements and acceptance}
For two-stream growth, the primary comparison is the ratio of mean individual rates,
\begin{equation}
 \widehat\gamma_s=\operatorname{slope}_{t\in I}\log|\widehat E_{x,s}(t)|,
 \qquad \overline\gamma=\frac1{N_s}\sum_s\widehat\gamma_s,
 \label{Equation16}
\end{equation}
where $\widehat E_{x,s}$ is the target-mode Fourier coefficient and the slope is an ordinary least-squares fit on a specified interval $I$. We also retain the original ensemble-amplitude observable, obtained by fitting $\log[N_s^{-1}\sum_s|\widehat E_{x,s}|]$. A fit to the average amplitude need not give the average of the individual fitted rates, because rapidly growing realisations contribute more strongly to the average at later times. We therefore report the two quantities separately. 

Paired 95\% intervals use 5,000 bootstrap resamples of seed identities, keeping each candidate and its CIC baseline together~\cite{efron1979bootstrap}. In our rxperiment crioteria, rate preservation requires the complete interval to lie in $[0.90,1.10]$. The CIC baseline and the filtered computation must likewise yield a signal that is long enough to be reliably measured: at minimum, two fitted amplitude e-foldings, a two-stream fit-start signal-to-floor ratio of at least five, and a perturbation-weight bound that matches the linear calculation. Supplementary Sections~S7 and S10 define these checks.

In the Landau damping case, the complex coefficients are first averaged, then used to build a geometric envelope, and the logarithm of that envelope is fitted. The resulting decay depends on both the damping of individual signals and their relative phases. In the projected Landau calculation the initial perturbation excites the forward Langmuir wave and a weaker backward wave with the same damping. There we fit two complex exponentials to the target-mode coefficient on $t\in[8,65]$ by the matrix-pencil method~\cite{hua1990matrix} and report the decay rate of the dominant forward root, together with the ordinary least-squares log-amplitude slope. Two-stream predictions include the mesh and midpoint-update time-step response.  To account for seed-to-seed variation, we perform separate calculations for each initial marker sample. The resulting response models are derived in Supplementary Section~S5.
\section{Results and discussion}
\label{section4}
We first examine how the same Gaussian filter affects different two-stream equilibria. We then apply the constrained selection jointly to Gaussian and compensated-binomial candidates, test the selected filters in projected-linear dynamics and explain the role of the filter shape and of finite-marker sampling.

\subsection{Two-stream response to Gaussian filtering}

\begin{figure*}[t]
\centering\includegraphics[width=\textwidth]{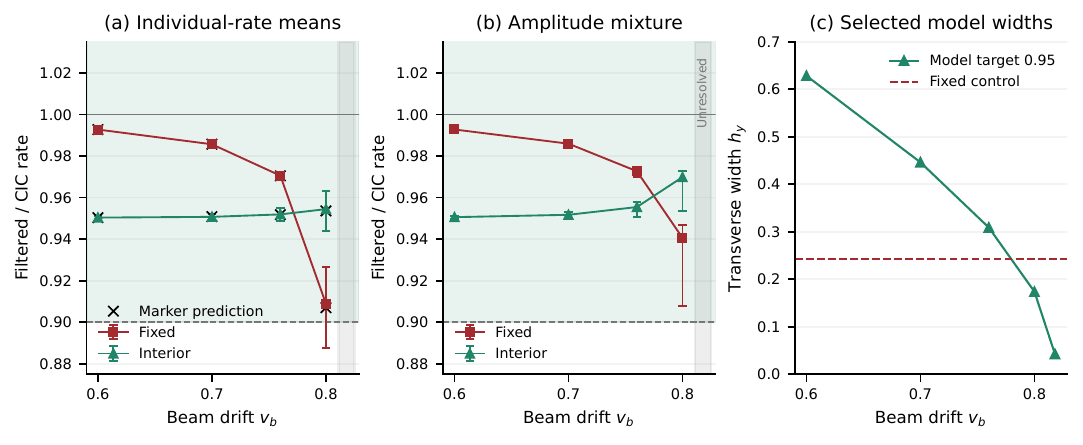}
\caption{Two-stream drift scan. (a) Ratios of mean individual growth rates; black crosses show seed-specific principal-mode predictions. The widths use the continuum-velocity grid/time constraint. (b) Ensemble-amplitude rate ratios. Red squares denote the fixed control and green triangles the interior settings. Error bars show paired 95\% confidence intervals and shading marks the tolerance band; the horizontal dashed line indicates its lower limit. (c) Interior widths and the fixed control (red dashed line). The grey strips mark the unresolved terminal drift in (a) and (b).}
\label{Figure2}
\end{figure*}

In Figure~\ref{Figure2}, we compare a fixed Gaussian with interior settings targeting $\Rg=0.95$ at each drift. We kept the initial spatial density and fixed-filter attenuation unchanged; then the fixed ratio of mean individual rates decreases from 0.99275 at $\vb=0.60$ to 0.90891 at $\vb=0.80$ (Figure~\ref{Figure2} (a)). The corresponding growth-rate reduction increases from 0.72\% to 9.11\%, demonstrating the equilibrium dependence of the sampled response. The change therefore comes from the equilibrium response and its finite-particle representation, rather than from a change in filter strength or initial density.

When the width is instead chosen separately at each drift to target $\Rg=0.95$, the measured ratios remain between 0.9504 and 0.9545, with all four confidence intervals inside the 10\% tolerance. Maintaining this response requires the transverse width to decrease from 0.6284 to 0.1739 cells (Figure~\ref{Figure2} (c)). The interior filter is broader than the fixed control at low drift and narrower at high drift, compensating for the increasing sensitivity to attenuation. At $\vb=0.80$ the fixed control lies, by construction, on the model's 10\% boundary. Its interval $[0.88743,0.92666]$ crosses the lower acceptance limit of 0.90, so the data establish neither preservation within 10\% nor a larger distortion. The ensemble-amplitude ratios lie increasingly above the individual-rate ratios as the drift grows (Figure~\ref{Figure2}(b)).
 
At $\vb=0.818$, just below the grid/time marginal drift 0.8192, the grid/time CIC rate is only 0.02928 $\omega_p$. The arm-level signal-to-floor ratios at the fit start are 2.9--3.6, below the required five, and three of the ten seeds in each arm exceed the perturbation-weight bound, so no ratio is accepted at this drift and the terminal width in Figure~\ref{Figure2}(c) remains a model setting. The fixed control's transfer, $W^2=0.99095$, lies below the model's marginal coupling 0.99715 at this drift, so the model has no growing root. Full values are given in Supplementary Sections~S2 and S8.

\subsection{Bandwidth selection and validation}

We next test whether the response constraint allows density-error reduction while preserving the protected mode's growth rate.  At $\vb=0.80$ the candidate set contains 115 filters: CIC, 84 Gaussians and the 30 compensated binomials. All are applied offline to the same CIC evolved particle states. The objective, Eq.~\eqref{Equation13} is averaged over five selection seeds. The constraint uses the grid/time model of the protected $(12,1)$ mode with $\varepsilon_{\rm op}=0.05$. Every Gaussian and 17 binomials have a growing root at their transfer. The other 13 binomials reach marginality on the tracked branch, and no root search is unresolved. Out of these candidates, 42 Gaussians and 11 binomials satisfy the 5\% limit. Table~\ref{Table2} compares three selection rules on this common sample.
 
\begin{table*}[t]
\centering\small
\caption{Selection rules on the common 115-candidate sample at $\vb=0.80$.}
\label{Table2}
\begin{tabularx}{\textwidth}{@{}lLrrrL@{}}\toprule
Selection rule & Selected filter & $W^2$ & Density MSE reduction & Field MSE reduction & Protected-mode growth\\\midrule
Density error only & Raw binomial, 8 passes & 0.13143 & 95.40\% [95.13, 95.65] & 85.27\% [84.17, 86.41] & Model: branch marginal at $W_c^2=0.95319$. Measured: no resolved growth\\\addlinespace
Constrained, Gaussian family & Gaussian $(0,0.1716)$ & 0.99547 & 7.16\% [6.99, 7.33] & 3.50\% [3.30, 3.72] & 0.95565 [0.94527, 0.96401] (0.95130)\\\addlinespace
Constrained, joint families & Compensated $(p,n)=(4,8)$ & 0.99585 & 76.44\% [75.05, 77.72] & 37.60\% [35.24, 40.13] & 0.95945 [0.94995, 0.96702] (0.95548)\\\bottomrule
\end{tabularx}
\end{table*}
 
\begin{figure*}[t]
\centering\includegraphics[width=\textwidth]{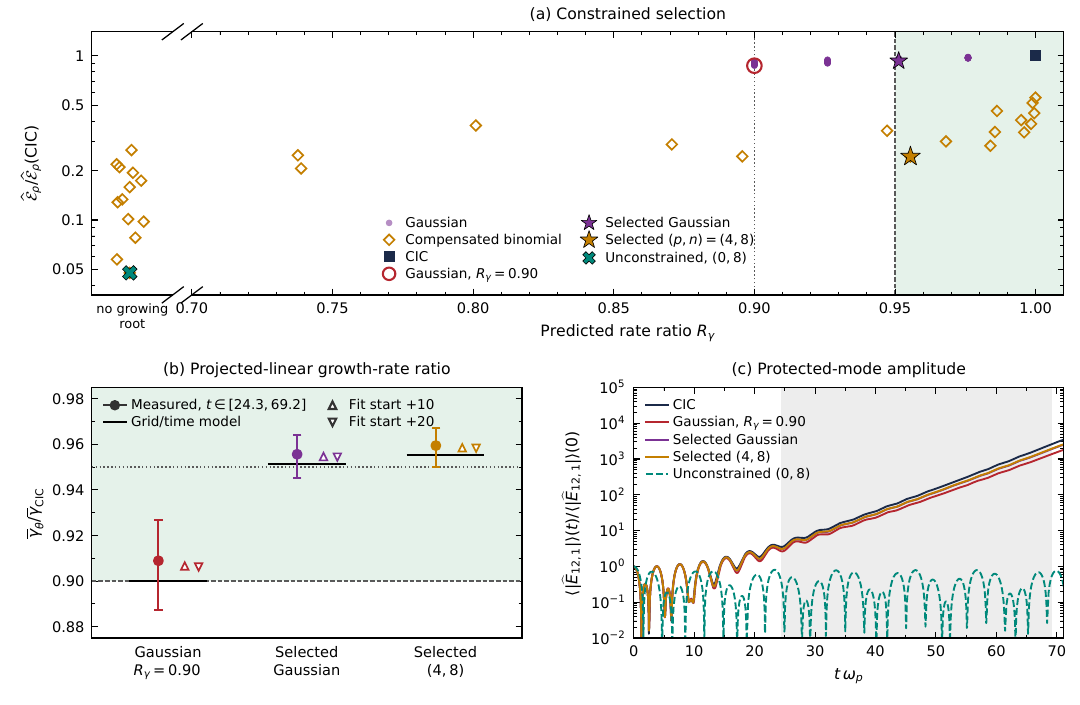}
\caption{Constrained selection and validation at $\vb=0.80$. (a) Density MSE on the selection seeds, relative to CIC, against the predicted growth-rate ratio for all 115 candidates. Shading marks the admissible region $\Rg\ge0.95$ (dashed line) and the dotted line the 10\% model boundary. (b) Ratios of mean individual growth rates on the primary fit window with paired 95\% seed intervals, grid/time predictions (black bars) and the ratios obtained when the fit start is delayed by 10 and 20 time units. Shading marks the 10\% acceptance band (lower limit dashed) and the dotted line the 0.95 operating target. (c) Ensemble amplitude of the protected mode relative to its initial value, the grey band is the primary fit window.}
\label{Figure3}
\end{figure*}

In Figure~\ref{Figure3}(a), without the constraint, the density objective selects the raw eight-pass binomial filter. Its round-trip transfer at the protected mode, $W^2=0.13143$, lies far below the marginal coupling $W_c^2=0.95319$, so the tracked branch has reached marginality. Across the fit window the CIC ensemble amplitude of the protected mode grows by a factor of 739, whereas the raw-filter amplitude changes by a factor of 1.23 and never exceeds its initial value (Figure~\ref{Figure3}(c)). Log-linear fits give a mean individual slope of 0.0025$\,\omega_p$, with $R^2<0.0013$ and 0.10-0.13 e-foldings in every seed, so no exponential growth is resolved. The largest reconstruction gain in the candidate set is therefore obtained by removing the protected growth, in the linear test as in the model. Whereas, with the constraint, the joint search selects the compensated binomial filter of degree $p=4$ with eight passes, $(p,n)=(4,8)$, whose model ratio is $R_{\gamma,\mathrm{model}}=0.95548$. On the evaluation seeds,
\begin{equation}
\begin{aligned}
 \mathcal N_\rho&=4.2442~[4.0081, 4.4886],\\
 \mathcal N_E&=1.6026~[1.5441, 1.6703],
\end{aligned}
 \label{Equation18b}
\end{equation}
that is, 76.44\% lower density MSE and 37.60\% lower field MSE than CIC. Restricting the same rule to the Gaussian family selects
\begin{equation}
 (h_x,h_y)_*=(0,0.1716393),\qquad R_{\gamma,\mathrm{model}}=0.95130,
 \label{Equation17}
\end{equation}
with $\mathcal N_\rho=1.0772~[1.0752, 1.0791]$ and $\mathcal N_E=1.0363~[1.0341, 1.0386]$, or reductions of 7.16\% and 3.50\%. Allowing the Gaussian to reach the 10\% model boundary raises its density-MSE reduction only to 13.24\%. Because $h_x=0$, the best admissible Gaussian preserves the initial longitudinal $(12,0)$ signal while smoothing transverse fluctuations. Within the tested families and candidate set, $(4,8)$ is the best admissible candidate for this objective. We then apply each selected filter throughout the projected-linear evolution of the protected mode, on the ten drift-scan seeds, which are disjoint from the selection and evaluation samples, each paired with the CIC run of the same seed. The ratios of mean individual rates on the primary window are
\begin{equation}
\begin{aligned}
 (p,n)=(4,8):\quad &\overline\gamma_*/\overline\gamma_\CIC=0.959452\;[0.949954, 0.967020],\\
 \text{Gaussian:}\quad &\overline\gamma_*/\overline\gamma_\CIC=0.955645\;[0.945273, 0.964006].
\end{aligned}
 \label{Equation19}
\end{equation}
Both intervals lie inside the 10\% acceptance range and contain the grid/time predictions, 0.95548 and 0.95130. All ten seeds satisfy the specified quality checks for both filters. Together, these two experiments over Gaussian and compensated-binomial filters finds a filter that lowers density and grid-field reconstruction errors, while still meeting the specified growth-rate tolerance. 
 
\subsection{Effect of filter design}

\begin{figure*}[t]
\centering\includegraphics[width=\textwidth]{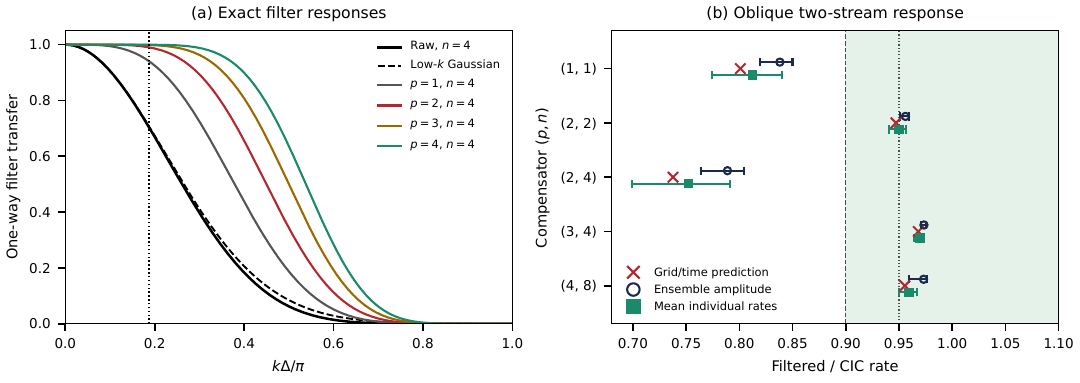}
\caption{Compensated-filter transfer and two-stream response. (a) One-way transfers for four binomial passes at different compensation degrees and the low-wavenumber Gaussian approximation. The vertical line marks the protected longitudinal wavenumber, $m_x=12$ on $N_x=128$. (b) Continuum-velocity grid/time predictions and both measured rate statistics, with paired 95\% intervals. Shading marks the 10\% tolerance band. The dotted line marks the 5\% operating limit.}
\label{Figure4}
\end{figure*}

The modest Gaussian error reduction is partly a consequence of its transfer function. Broadening a Gaussian to suppress more short-wavelength noise also increases attenuation at the protected mode. Compensation changes this balance by keeping the transfer closer to unity over a wider range of low wavenumbers. Gaussian attenuation begins at quadratic order in wavenumber, whereas order-$p$ compensation delays the leading departure from unity to order $(k\Delta)^{2p+2}$, and every binomial transfer vanishes at the Nyquist wavenumber. Figure~\ref{Figure4}(a) shows how this flatter passband retains the protected mode while attenuating shorter wavelengths.

The CIC error is spread over all resolved wavelengths. For a representative seed, only 0.42\% of its power lies in the protected mode and 43\% at $|k_x|$ above half the Nyquist wavenumber. A filter that is nearly transparent at the protected mode can therefore still remove most of the error, provided its transfer falls steeply at shorter wavelengths. The selected $(4,8)$ filter attenuates the protected mode slightly more than $(3,4)$ ($W^2=0.99585$ against 0.99700) but suppresses grid-scale fluctuations more strongly, so the density objective prefers it.
 
The four representative filters in Figure~\ref{Figure4}(b) show why compensation degree alone is insufficient. The $(3,4)$ filter has model ratio 0.968 and measured ratio 0.96977 [0.96436, 0.97376], satisfying both the 5\% model operating limit and the 10\% measured tolerance, and reduces the density MSE by 70.5\% on the common sample. The $(2,2)$ setting passes the 10\% measured tolerance with ratio 0.94996 [0.94092, 0.95664], but its model ratio 0.94717 fails the stricter 5\% operating constraint. Increasing the pass count to $(2,4)$ moves the response outside the 10\% band, and $(1,1)$ also fails. These results show that the filter must also suppress fluctuations where they contribute to the error while leaving enough coupling at the protected mode.

\subsection{Finite-particle effects on measured growth}
\label{sec:finite-marker}
The growth rates differ between seeds even though the prescribed equilibrium and filter are the same. At $v_b=0.80$, the CIC rates have mean $0.124451\,\omega_p$ and standard deviation $0.023328\,\omega_p$. Each run represents the velocity distribution with a finite set of markers, so different realisations need not reproduce exactly the same response~\cite{touati2022ensemble}.

\begin{figure*}[t]
\centering\includegraphics[width=\textwidth]{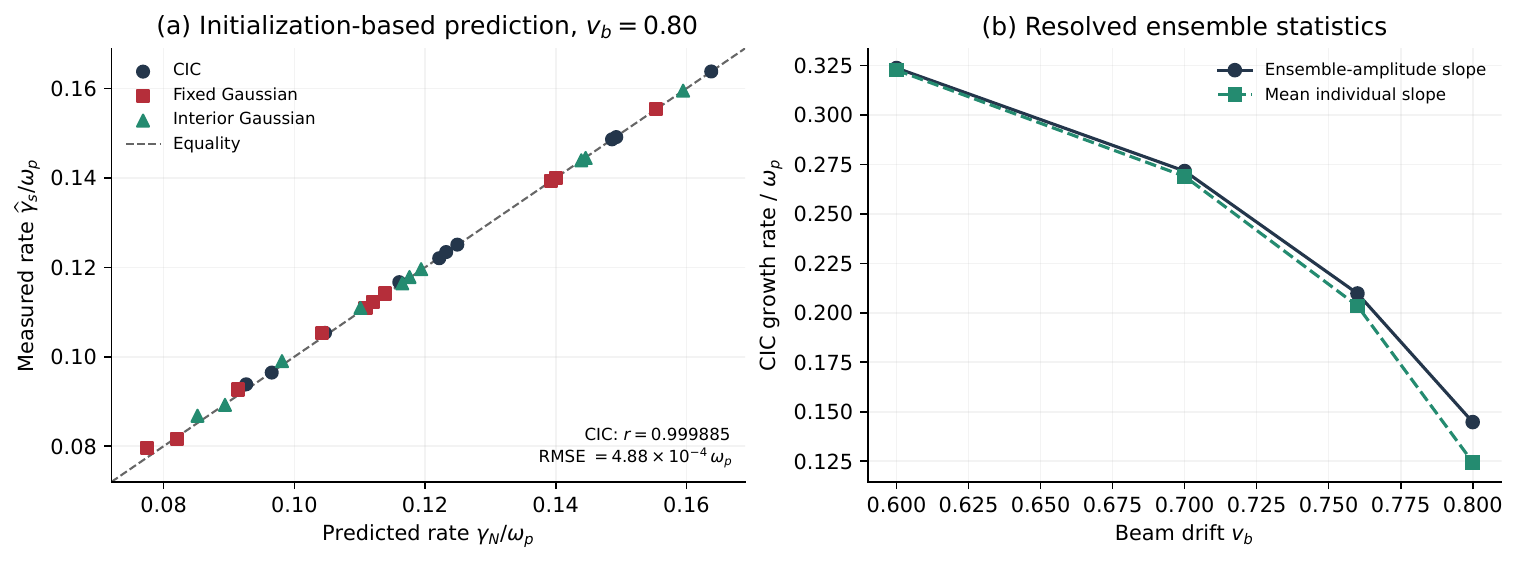}
\caption{Finite-marker prediction and estimator dependence. (a) Measured individual growth rates versus initialization-based principal-mode predictions at $\vb=0.80$, for CIC, the fixed Gaussian and the interior Gaussian (ten seeds each). The dashed line is equality. (b) CIC ensemble-amplitude slopes and mean individual rates across the four resolved drifts, using the same ten seeds and prescribed window at each drift.}
\label{Figure5}
\end{figure*}

To test whether this sampling explains the observed spread, we evaluate a principal-mode response using the reconstructed initial velocities and the linearised source $\nabla_\vv\log f_0$. No fitted growth rates enter this prediction. It closely follows the measured seed variation: the CIC correlation is $r=0.999885$ and the RMSE is $4.88\times10^{-4}\,\omega_p$ (Figure~\ref{Figure5}(a)). Predicted fixed/CIC and interior/CIC ratios are 0.90702 and 0.95360, compared with measured values 0.90891 and 0.95446. The seed variation is therefore largely explained by the sampled velocity response, without using measured rates as predictor inputs. The derivation and approximation limits are given in Supplementary Sections~S5 and S8.

Ensemble averaging adds a second effect. For exponential seed signals,
\begin{equation}
 \frac{d}{dt}\log\sum_s A_s e^{\gamma_s t}
 =\frac{\sum_s\gamma_s A_s e^{\gamma_s t}}{\sum_s A_s e^{\gamma_s t}}.
 \label{Equation20}
\end{equation}
Faster-growing seeds acquire more weight, so an ensemble-amplitude fit generally exceeds the mean individual rate. At $v_b=0.80$, the two CIC estimates are $0.144746\,\omega_p$ and $0.124451\,\omega_p$. Reconstructing the amplitude mixture from the individual fits gives $0.144731\,\omega_p$, accounting for the difference seen in Figure~\ref{Figure5}(b). This explains why the two estimators can give different acceptance decisions near the tolerance boundary. We use the ratio of mean individual rates for the primary growth comparison and retain the ensemble observable as a distinct measurement.

\subsection{Weak Landau damping}

\begin{figure*}[t]
\centering\includegraphics[width=\textwidth]{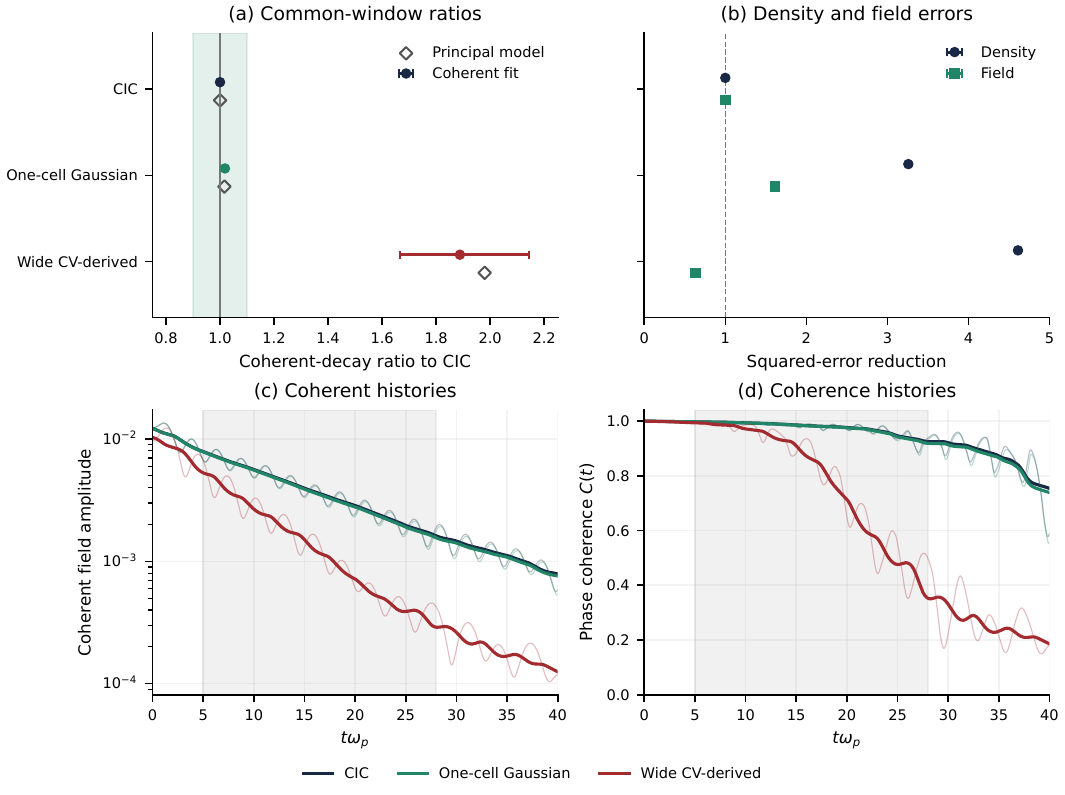}
\caption{Weak-Landau comparison using a common fitting window $[5,28]$. (a) Measured coherent-decay ratios and continuous-time principal-harmonic predictions; shading marks the 10\% tolerance. (b) Density and field squared-error reduction factors. (c) Coherently averaged target-mode amplitudes (thin curves) and their 49-sample geometric envelopes (thick curves). (d) Amplitude-weighted phase coherence and its geometric envelope. Shading in (c,d) marks the common fit interval.}
\label{Figure6}
\end{figure*}

We next examine whether improved density estimation also reduces field error and preserves the decay of a weakly damped mode. Using 40 paired seeds, we compare a one-cell Gaussian filter with a wider filter selected in an earlier cross-validation study. Over the common fitting interval $[5,28]$, their coherent-ensemble decay-rate ratios relative to CIC are 1.01814 [1.01486, 1.02121] and 1.88805 [1.66753, 2.14374], respectively. The one-cell interval lies entirely within the 10\% rate tolerance, whereas the wider filter substantially increases the measured decay rate. In Figure~\ref{Figure6}(a), we compare these measurements with the principal-harmonic predictions.

The wider filter reduces density error by a factor of 4.612 but increases the field error by approximately $1.58\times$. In contrast, the one-cell filter reduces both errors, by factors of 3.259 and 1.613, respectively (Figure~\ref{Figure6}(b)). Thus, a larger improvement in density estimation need not improve field accuracy or preserve coherent decay.

CIC, the one-cell filter and the wide comparator cover 1.582, 1.611 and 2.987 amplitude e-foldings, respectively, on the common window. The baseline therefore fails the adopted two-e-folding requirement for both comparisons. The one-cell filter satisfies the coherent-ratio criterion and improves both error metrics, but it does not achieve complete confirmation under the stated protocol.

The coherent average can decrease for two reasons: individual signals can damp, and their phases can separate so that they cancel more strongly when added. Let $M=N_s^{-1}\sum_s|\widehat E_s|$ be the mean magnitude and $C=|\sum_s\widehat E_s|/\sum_s|\widehat E_s|$ measure their phase coherence. The coherent amplitude is then $A=MC$. Taking logarithms and applying the same envelope and slope fit gives
\begin{equation}
 \widehat\gamma_{\rm coh}=\widehat\gamma_M+\widehat\gamma_C.
 \label{Equation21}
\end{equation}
For the wide filter, the coherence term accounts algebraically for 63.8\% of the wide-minus-CIC slope difference (Supplementary Section~S9.1). The corresponding mean-magnitude ratio is 1.335, compared with the coherent-decay ratio 1.888. Neither term alone estimates the damping eigenvalue: the mean-magnitude slope is amplitude-weighted towards slower-decaying seeds, while dephasing contributes a negative coherence slope (Figure~\ref{Figure6}(d)). The principal-harmonic predictions, 1.016 and 1.980, lie inside the one-cell and wide coherent-ratio intervals. The agreement is therefore consistent with the response model, but a coherent decay rate alone cannot identify the underlying damping eigenvalue. In particular, the 63.8\% contribution describes how the measured slope is composed. It does not assign that fraction of an eigenvalue change to dephasing.

The finite reference also limits the interpretation of density rankings. Its variance contributes 19.5\% of the CIC density MSE. If candidate-reference covariance is negligible, then it implies an approximate maximum reduction factor of $1/0.195\simeq5.1$. The wide Gaussian comparator's factor 4.612 already reaches about 90\% of that ceiling. Therefore, its apparent density benefit is observed only close to the reference noise floor and does not compensate for the worsening of field error or coherent response. Supplementary Section~S9.2 provides the complete bias–variance–covariance breakdown.

\subsection{Findings and limitations}
This study implies that a filter must be chosen for the response one wants to retain. The density objective measures how well a particle sample represents spatial structure. Meanwhile, the dispersion relation indicates how variations in the particle–field coupling affect the growth or damping of a mode. Putting the two together yields a well-justified selection of smoothing intensity. The Gaussian study illustrates this decision, and the compensated-filter outcome indicates that refining the transfer response can boost the statistical gain achievable under the same rate constraint.

These tests focus on fixed, spatially uniform filters and on specified linear modes. In the two-stream case, the thermal scale $\vth/\omega_p=0.05$ is smaller than either grid spacing.Reconstruction errors are measured on offline-filtered CIC snapshots, whereas growth rates are measured in separate projected-linear calculations. The reported error reductions therefore do not establish the same improvement during self-consistently filtered nonlinear evolution.

For a time-dependent plasma, using this approach would also mean deciding when to refresh the equilibrium model and which modes to keep. Filters that vary in space would require accounting for their nonuniform coupling and conservation behavior. The results presented here address the simpler fixed-filter setting to show a permissible modification of the linear dynamics into a practical constraint on statistical smoothing.

\section{Conclusions}
\label{Section5}
We introduced a bandwidth-selection approach that incorporates how smoothing influences linear plasma behaviour. Using the dispersion relation, the method finds filters that preserve the growth or damping rates of chosen modes with the targeted tolerance, and then picks among these candidates by minimising the density error. A controlled two-stream parameter scan shows why the equilibrium must be part of this decision: with the same Gaussian transfer, growth-rate reductions increases by more than an order of magnitude as the beam drift approaches marginal stability, as the model predicts.

In the Gaussian candidate search, the selected width reduces density and grid-field MSE by 7.16\% and 3.50\% on independent evaluation samples. Used in the separate projected-linear computation, it yields a growth-rate ratio of 0.9556, with its paired interval remaining within the 10\% tolerance. A compensated $(4,8)$ binomial filter preserves the desired rate with greater margin and reduces density MSE by 76.4\%. Overall, these findings indicate that selecting the transfer shape, not only its magnitude can lead to much larger gains while still satisfying a dynamical constraint.

The Landau comparison shows the importance of checking the field and the measured decay alongside density error. The widest filter gives the largest density improvement but worsens field accuracy and changes coherent decay. The short CIC baseline constrains how robust the rate evaluation can be. Within the linear configurations tested here, the proposed constraint makes the intended plasma response an explicit part of smoothing selection. 

\section*{Data availability}
The research code, archived simulation histories, configurations, reproduction and analysis scripts are available in the Github repository, cited as Agrawal et al.~\cite{agrawal2026code}.

\section*{Declaration of competing interest}
The authors declare no conflicts of interest.

\section{CRediT authorship contribution statement}
\textbf{A. Agrawal}: Conceptualization, Methodology, Formal analysis, Validation, Writing - original draft; \textbf{S. C. Lin}: Formal analysis, Validation, Writing – review \& editing; \textbf{E. Krishnasamy:} Software, Formal analysis, Validation, Writing – review \& editing, Supervision, Funding acquisition, Project administration; \textbf{P. Bouvry:} Project administration.

\section*{Acknowledgements}
This work was supported by the Fonds National de la Recherche (FNR), Luxembourg, under the \textbf{FNR CORE} project grant number \textbf{{C23/IS/18105668/HEXAPIC}}. The authors express their gratitude to EuroHPC JU for computing resources, specifically the MeluXina, LUMI, and MareNostrum 5 systems, which were utilized for the experimental tests conducted in this study. Additionally, we thank the LRZ SuperMUC-NG Phase 2 for granting access to Intel GPUs.  
 
\FloatBarrier

\appendix
\setcounter{figure}{0}
\setcounter{table}{0}
\renewcommand{\thefigure}{A\arabic{figure}}
\renewcommand{\thetable}{A\arabic{table}}
\section{Linear susceptibility and reference rates}
\label{A1}
The susceptibilities follow from the linearised Vlasov--Poisson system, using the time convention in Section~\ref{Section2}. For a unit-normalised equilibrium velocity density $F_0$ and total electron plasma frequency $\omega_p$, the electrostatic susceptibility is~\cite{fried1961dispersion,adams2023petscpic}
\begin{equation}
 \chi(\kk,\omega)=\frac{\omega_p^2}{|\kk|^2}
 \int\frac{\kk\cdot\nabla_{\vv}F_0(\vv)}{\omega-\kk\cdot\vv}\,d^2v.
 \label{eq:susceptibility-integral}
\end{equation}
The integral is initially defined for $\operatorname{Im}\omega>0$ and analytically continued using the Landau prescription. Integrating over the velocity perpendicular to $\kk$ reduces each isotropic Maxwellian contribution to a one-dimensional Gaussian integral. The plasma dispersion function is
\begin{equation}
 Z(\zeta)=\frac1{\sqrt\pi}\int_{-\infty}^{\infty}
 \frac{e^{-u^2}}{u-\zeta}\,du,\qquad\operatorname{Im}\zeta>0,
 \label{eq:plasma-Z}
\end{equation}
with analytic continuation elsewhere~\cite{fried1961dispersion}. For a Maxwellian with thermal standard deviation $\vth$ and $\lambda_D=\vth/\omega_p$, the principal susceptibility is
\begin{equation}
 \chi_L(k,\omega)=\frac{1}{k^2\lambda_D^2}[1+\zeta Z(\zeta)],
 \qquad \zeta=\frac{\omega}{\sqrt2k\vth},
\end{equation}
Here $k=|\kk|$; substituting the Maxwellian derivative in Eq.~\eqref{eq:susceptibility-integral} and using $u/(u-\zeta)=1+\zeta/(u-\zeta)$ gives this form. Linearity of the velocity integral then gives, for two equally weighted isotropic Maxwellian beams,
\begin{equation}
 \chi_{\rm E2}(\kk,\omega)=\frac{\omega_p^2}{2|\kk|^2\vth^2}
 \sum_\pm[1+\zeta_\pm Z(\zeta_\pm)],\qquad
 \zeta_\pm=\frac{\omega\mp k_x\vb}{\sqrt2|\kk|\vth}.
\end{equation}
The subscript E2 identifies the oblique two-stream benchmark defined in Section~\ref{Section3}. Each beam contributes half the total electron density. The thermal and Poisson factors use $|\kk|$, while the drift resonance uses $k_x\vb$. Roots are continued from the corresponding CIC branch as attenuation increases. The thermal scale $\lambda_D=\vth/\omega_p$ uses the total electron density; it does not replace the full drift-dependent susceptibility of the two-beam equilibrium.

At $\vb=0.80$, the archived longitudinal two-stream principal-CIC rates are $\gamma/\omega_p=0.34697$ at $m_x=8$ and 0.14467 at $m_x=12$, compared with grid/time values 0.346960 and 0.144600. At the oblique corner, the principal transverse 10\% boundary is approximately 0.242831 cells and the numerical boundary is 0.242735 cells. The curves in Figure~\ref{Figure1} use the principal continuous-time response. The Gaussian operating settings and compensated-filter predictions use the continuum-velocity grid/time response derived in Supplementary Section~S5.2. The seed-specific predictor in Supplementary Section~S5.3 is a separate principal-mode approximation.

For Landau, the continuum root is $\omega/\omega_p=1.285057-0.066128i$. With the principal CIC shape, continuous time gives $\gamma/\omega_p=-0.06630795$. At $\Delta t=0.05$, the standard velocity-Verlet position update gives a homogeneous principal-CIC reference of approximately $-0.0662846$. These values illustrate the distinction between continuous-time and integrator-specific rates~\cite{langdon1979time}. Both retain continuum velocity integration and omit spatial aliases. Supplementary Section~S5.1 derives the discrete reference and distinguishes it from the midpoint-weight value in the historical archive. Paired measured rates use the same update within each campaign.

At one nonzero wavenumber for which $0<|\cos(k\Delta/2)|\le1$, set the raw-binomial transfer $\cos^{2n}(k\Delta/2)$ equal to $\exp[-k^2\sigma_{\rm eq}^2/2]$ and take logarithms. This gives the exact mode-specific Gaussian equivalent
\begin{equation}
 \sigma_{\rm eq}^2(k)=-\frac{4n}{k^2}\log|\cos(k\Delta/2)|.
\end{equation}
Its low-$k$ limit is $n\Delta^2/2$. Since the exact width depends on $k$, a single Gaussian generally does not reproduce the binomial response at all protected modes.

\section{Charge, force and energy checks}
\label{app:conservation}
Here the particle charge $Q_p$ and field $\EE_p$ are those in Eq.~\eqref{Equation4}. In a 4096-particle deposition--solve--gather test, the normalised internal-force sum $\|\sum_pQ_p\EE_p\|_2/\allowbreak\sum_p\|Q_p\EE_p\|_2$ is $1.37\times10^{-17}$, $1.00\times10^{-17}$ and $1.06\times10^{-18}$ at scalar cell widths 0, 0.5 and 3. Represented charge is constant in all 90 archived spectral energy histories. These checks support the fixed matched operator in Eq.~\eqref{Equation4}.

Charge and force conservation do not ensure exact energy conservation for the explicit full-$f$ update~\cite{brackbill2016conservation}. The separate conservation campaign reports maximum relative errors $|\mathcal E_{\rm tot}(t)-\mathcal E_{\rm tot}(0)|/|\mathcal E_{\rm tot}(0)|$ in total kinetic-plus-electrostatic energy of $4.93\times10^{-6}$ for CIC and $3.30\times10^{-7}$ for the filtered weak-Landau configuration; the two-stream values are $6.73\times10^{-3}$ and $4.97\times10^{-3}$. Eight of nine seed-bootstrap energy-slope intervals exclude zero. These records concern the full-$f$ energy experiments, not the fixed-velocity projected scan.

\FloatBarrier
\bibliographystyle{elsarticle-num}
\bibliography{references}
\end{document}


\begin{frontmatter}
\title{Supplementary material for\\ ``Dynamics-aware bandwidth selection for smoothed charge deposition in 2D electrostatic particle-in-cell''}
\author[lux]{Ayushi Agrawal}
\ead{ayushi.agrawal@uni.lu}
\author[lux]{Shan-Chang Lin}
\author[lux,lj,rud]{Ezhilmathi Krishnasamy}
\author[lux]{Pascal Bouvry}
\cortext[cor1]{Corresponding author.}
\address[lux]{Faculty of Science, Technology and Medicine, University of Luxembourg, Esch-sur-Alzette, Luxembourg}
\address[lj]{University of Ljubljana, Ljubljana, Slovenia}
\address[rud]{Rudolfovo Science and Technology, Novo Mesto, Slovenia}
\begin{abstract}
This document supplements the main article. Sections, equations, figures, tables and pages are numbered with the prefix S; numbers without the prefix (for example, Eq.~(8) or Fig.~3) refer to the main text. 
\end{abstract}
\end{frontmatter}

\section{Overview, conventions and terminology}
\label{sec:overview}

This supplement states the complete numerical comparisons behind Section~4 of the main text. Table~\ref{tab:roadmap} shows how the sections connect to the main text.

\begin{table}[!htbp]
\centering\small
\caption{Organisation of the supplement and its connection to the main text.}
\label{tab:roadmap}
\begin{tabular}{@{}lP{0.52\linewidth}P{0.26\linewidth}@{}}\toprule
Section & Content & Main text\\\midrule
\ref{sec:designs} & Numerical configurations, sample assignments and complete comparisons & Sections~3.1-3.2, 4.1-4.3, 4.5\\
\ref{sec:loading}-\ref{sec:coupling} & Particle loading, initial perturbation, particle-field coupling and $\delta f$ update & Section~3.1\\
\ref{sec:response} & Response models: principal harmonic, grid/time, finite marker; root status and marginal coupling & Sections~2.2, 2.4, 3.3, 4.4; Appendix~A\\
\ref{sec:selection} & Reconstruction-error objective, constrained selection & Sections~2.3, 3.2, 4.2\\
\ref{sec:estimators} & Rate estimators, uncertainty, acceptance checks and fit-window sensitivity & Sections~3.3, 4.1-4.2\\
\ref{sec:drift-results} & Drift-scan measurements and finite-marker predictions & Sections~4.1, 4.4\\
\ref{sec:landau} & Landau envelopes, phase coherence and finite-reference error & Section~4.5\\
\ref{sec:exact-width} & Dynamics test of the selected filters and the unconstrained optimum & Section~4.2\\
\ref{sec:landau-df} & Projected-linear Landau damping & Section~4.5\\\bottomrule
\end{tabular}
\end{table}

\paragraph{Normalisation}
All calculations are electrostatic and use
\begin{equation}
 \omega_p=1,\qquad n_0=1,\qquad \epsilon_0=1,\qquad m_e=1,\qquad q_e=-1,
\end{equation}
where $n_0$ is the equilibrium electron number density and $\omega_p^2=n_0q_e^2/(m_e\epsilon_0)$. The Debye length is $\lambda_D=\vth/\omega_p$. It equals 1 in the Landau calculation and 0.05 in the two-stream calculations, where it is smaller than both grid spacings ($\Delta x=0.5$, $\Delta y=1$). The electrons obey the periodic Vlasov-Poisson system~\cite{birdsall2004plasma,hockney1988computer}
\begin{equation}
 \partial_t f+\vv\cdot\nabla_{\xx}f-\EE\cdot\nabla_{\vv}f=0,\qquad
 -\nabla^2\phi=\rho,\qquad \EE=-\nabla\phi,
 \label{eq:vlasov}
\end{equation}
with an immobile ion background that cancels the uniform equilibrium electron charge. Filter widths are written in cell units, $h_i=\sigma_i/\Delta_i$, as in the main text.

 \textbf{Terminology}

\begin{itemize}
\item \emph{Protected mode}: the oblique two-stream mode $(m_x,m_y)=(12,1)$, whose growth rate the constraint preserves.
\item \emph{Fixed control} (also \emph{boundary Gaussian}): the Gaussian $(h_x,h_y)=(0,0.2427346)$, applied unchanged at every beam drift. At $\vb=0.80$ it lies on the 10\% boundary of the grid/time model.
\item \emph{Interior setting}: a Gaussian with $h_x=0$, recomputed at each drift so that the grid/time model predicts $\Rg=0.95$.
\item \emph{Selected Gaussian}: the admissible Gaussian with the smallest density error, $(h_x,h_y)=(0,0.1716393)$. \emph{Selected binomial}: the admissible filter with the smallest density error over all candidates, the compensated binomial $(p,n)=(4,8)$. \emph{Unconstrained optimum}: the candidate with the smallest density error when the response constraint is omitted, the raw eight-pass binomial $(p,n)=(0,8)$.
\item \emph{Root status}: \emph{growing} if the branch continued from CIC reaches the candidate transfer with positive growth, \emph{marginal} if the tracked growth rate reaches zero at a coupling above the candidate transfer, and \emph{unresolved} if the continuation fails (Section~\ref{sec:root-status}). Only a growing root defines $\Rg$.
\item \emph{One-cell} and \emph{wide} Gaussians: the two filtered arms of the weak-Landau comparison, $(1,1)$ and $(7.2443,7.2443)$.
\item \emph{Seed}: one independent particle realisation. CIC and filtered runs are \emph{paired} when they start from the same seed.
\item \emph{Mean individual rate} $\overline\gamma$: the average of the rates fitted to each seed separately; the primary two-stream observable. \emph{Ensemble-amplitude rate} $\widehat\gamma_{\rm amp}$: the rate fitted to the seed-averaged amplitude (Section~\ref{sec:estimators}).
\end{itemize}

\section{Numerical configurations and complete comparisons}
\label{sec:designs}

Six sets of calculations supply the measurements of the main paper (Table~\ref{tab:configs}). Three measure two-stream growth rates in projected-linear runs with a filter active throughout the evolution. One measures reconstruction error: there the filters are applied to the deposits of common CIC-evolved particle states and do not act on the particles. The weak-Landau full-$f$ calculation measures both errors and damping in one simulation, and the projected Landau calculation measures damping rates over a longer baseline. Unless stated otherwise, brackets give percentile 95\% intervals from 5,000 paired bootstrap resamples of seeds (Section~\ref{sec:estimators}). Time samples within one simulation are correlated and are not treated as independent replicates.

\begin{table}[!htbp]
\centering\footnotesize
\caption{Numerical configurations. }
\label{tab:configs}
\begin{tabularx}{\linewidth}{@{}P{0.15\linewidth}P{0.19\linewidth}P{0.15\linewidth}P{0.13\linewidth}LP{0.13\linewidth}@{}}\toprule
Calculation & Evolution and filter use & Mode and amplitude & PPC; $\Delta t$ & Seeds & Measurement window\\\midrule
Drift scan & Projected linear $\delta f$; fixed control or interior setting active & $(12,1)$; $\alpha=10^{-8}$ & 512; 0.02 & 10 at each of $\vb=0.60$, 0.70, 0.76, 0.80, 0.818 & Fit from $t=24.32$; end in Table~\ref{tab:drift-design}\\\addlinespace
Exact-width test & As drift scan; selected Gaussian, selected $(4,8)$ binomial or raw $(0,8)$ binomial active & $(12,1)$; $10^{-8}$ & 512; 0.02 & The 10 drift-scan seeds at $\vb=0.80$, paired with their CIC runs & $[24.32,69.16]$\\\addlinespace
Reconstruction & Full $f$ with CIC coupling and random beam assignment; CIC, 84 Gaussian and 30 binomial candidates applied offline & $(12,0)$; $10^{-3}$ & 32 (reference 1024); 0.05 & 5 for selection; 5 disjoint for evaluation & Error over $t\in[0.5,4]$\\\addlinespace
Binomial rates & Projected linear $\delta f$; $(p,n)=(1,1)$, (2,2), (2,4), (3,4) in both directions & $(12,1)$; $10^{-8}$ & 512; 0.02 & 10 at $\vb=0.80$, a separate ensemble with its own paired CIC runs & $[24.32,59.68]$; evolved to $t=60$\\\addlinespace
Weak Landau & Full $f$, velocity Verlet; filter active in each arm; $64\times16$ mesh on $(5\pi,8)$, $\vth=1$ & $(1,0)$, $k\lambda_D=0.4$; $\alpha=0.01$ & 1024 (reference 4096); 0.05 & 40 paired seeds & Decay fit $[5,28]$; error over $[5,30]$\\\addlinespace
Landau response & Projected linear $\delta f$, explicit midpoint; CIC, one-cell and wide Gaussians active & $(1,0)$; travelling-wave initial weights & 1024; 0.05 & 3 paired seeds & Two-root fit $[8,65]$; slopes on several windows\\\bottomrule
\end{tabularx}
\end{table}

\subsection{Controlled two-stream drift scan}
\label{sec:drift-design}

The drift scan isolates the dependence of the filtered response on the equilibrium. The equilibrium is the equal two-beam mixture
\begin{equation}
 f_0(v_x,v_y)=\tfrac12 M_{\vth}(v_x-\vb)M_{\vth}(v_y)
              +\tfrac12 M_{\vth}(v_x+\vb)M_{\vth}(v_y),
 \qquad
 M_{\vth}(u)=\frac{\exp[-u^2/(2\vth^2)]}{\sqrt{2\pi}\,\vth}.
 \label{eq:f0}
\end{equation}
Within each seed, marker positions, thermal velocity samples, beam assignments and the initial perturbation are identical at every drift; only the symmetric beam offsets $\pm\vb$ change (Section~\ref{sec:loading}). The initial density, and therefore any density-based bandwidth objective evaluated on it, is the same at every drift, whereas the susceptibility is not.

Each drift compares CIC with two Gaussian settings. The fixed control has round-trip transfer $W_H^2=0.99095494$ at the protected mode for every drift, so the filter attenuation does not change across the scan. The interior setting is recomputed at each drift from the grid/time model of Section~\ref{sec:gridtime} so that the model predicts $\Rg=0.95$ with $h_x=0$. These widths are computed from the model before the filtered simulations are run (Table~\ref{tab:drift-design}). Fits start at $t=24.32$ at every drift and end at prescribed times that lengthen as the growth rate decreases.

\begin{table}[!htbp]
\centering\small
\caption{Interior widths and fit intervals for the drift scan. }
\label{tab:drift-design}
\begin{tabular}{@{}lrrr@{}}\toprule
$\vb$ & Interior $h_y$ & Fit start & Fit end\\\midrule
0.60 & 0.6284 & 24.32 & 40.76\\
0.70 & 0.4462 & 24.32 & 44.12\\
0.76 & 0.3088 & 24.32 & 50.80\\
0.80 & 0.1739 & 24.32 & 69.16\\
0.818 & 0.0425 & 24.32 & 204.08\\\bottomrule
\end{tabular}
\end{table}

At the terminal drift $\vb=0.818$ the equilibrium is close to marginal stability and the mode grows too slowly to be measured reliably. The arm-level signal-to-floor ratios at the start of the fit are 3.56 (CIC), 2.85 (fixed) and 3.53 (interior), below the required factor of five, and three of the ten seeds in each arm exceed the perturbation-weight bound of Section~\ref{sec:checks}. No rate is accepted at this drift, and the interior width $h_y=0.0425$ is a model setting without measured confirmation. The fixed control's transfer lies below the marginal coupling of this drift (Section~\ref{sec:root-status}), so the model has no growing root for it. The complete measured ratios are listed in Section~\ref{sec:drift-results}.

\subsection{Compensated-binomial comparison}
\label{sec:binomial-design}

This comparison asks whether the response constraint can guide the shape of the filter as well as its strength. The candidate family contains the 30 filters $(p,n)$ with compensation degree $p=0,\dots,4$ (degree 0 is the raw binomial) and pass count $n=1,2,3,4,6,8$, applied in both directions with the transfer of Eq.~\eqref{eq:binomial}. Their density and field errors are evaluated on the same reconstruction states and seeds as the Gaussian candidates (Section~\ref{sec:selection}). Growth rates are measured in projected-linear $\delta f$ runs with the drift-scan configuration at $\vb=0.80$: for the selected $(4,8)$ filter and the unconstrained optimum $(0,8)$ on the ten drift-scan seeds and fit window of the CIC baseline, and for four representative filters on a separate ten-seed ensemble with its own paired CIC runs, evolved to $t=60$ and fitted on $[24.32,59.68]$, over which the CIC baseline spans 4.48 e-foldings. Table~\ref{tab:binomial-complete} collects the results.

\begin{table}[!htbp]
\centering\footnotesize
\caption{Compensated-binomial and Gaussian settings at $\vb=0.80$. $W^2_{(12,1)}$ is the exact round-trip filter transfer at the protected mode, excluding the CIC factor. Model ratios use the continuum-velocity grid/time response (Section~\ref{sec:gridtime}); ``marginal'' means that the tracked branch has reached zero growth at a coupling above the filter transfer (Section~\ref{sec:root-status}), so no ratio is defined. Measured ratios are filtered/CIC with paired 95\% intervals. The error factors are $\mathcal N=\widehat{\mathcal E}(\CIC)/\widehat{\mathcal E}(\theta)$ on the common evaluation seeds of Section~\ref{sec:selection}. Rates for $(1,1)$-$(3,4)$ come from a separate ten-seed ensemble; all other rates use the drift-scan seeds.}
\label{tab:binomial-complete}
\setlength{\tabcolsep}{4pt}
\begin{tabular}{@{}lrrrrrr@{}}\toprule
Filter & $W^2_{(12,1)}$ & Model $\Rg$ & Mean individual rates & Ensemble amplitude & Density $\mathcal N_\rho$ & Field $\mathcal N_E$\\\midrule
$(p,n)=(1,1)$ & 0.98300 & 0.80096 & \vi{0.81224}{0.77428, 0.83989} & \vi{0.83807}{0.81930, 0.84994} & \vi{2.684}{2.616, 2.761} & \vi{1.348}{1.323, 1.377}\\
$(2,2)$ & 0.99509 & 0.94717 & \vi{0.94996}{0.94092, 0.95664} & \vi{0.95608}{0.95120, 0.95920} & \vi{2.901}{2.812, 2.997} & \vi{1.364}{1.335, 1.397}\\
$(2,4)$ & 0.97842 & 0.73769 & \vi{0.75240}{0.69884, 0.79128} & \vi{0.78885}{0.76359, 0.80469} & \vi{4.158}{3.950, 4.373} & \vi{1.613}{1.558, 1.676}\\
$(3,4)$ & 0.99700 & 0.96811 & \vi{0.96977}{0.96436, 0.97376} & \vi{0.97342}{0.97047, 0.97529} & \vi{3.393}{3.259, 3.539} & \vi{1.446}{1.406, 1.491}\\
$(4,8)$, selected & 0.99585 & 0.95548 & \vi{0.95945}{0.94995, 0.96702} & \vi{0.97312}{0.95912, 0.97595} & \vi{4.244}{4.008, 4.489} & \vi{1.603}{1.544, 1.670}\\
$(0,8)$, unconstrained & 0.13143 & marginal & no resolved growth & no resolved growth & \vi{21.727}{20.523, 22.999} & \vi{6.789}{6.318, 7.356}\\
\addlinespace
Selected Gaussian & 0.99547 & 0.95130 & \vi{0.95565}{0.94527, 0.96401} & \vi{0.97063}{0.95536, 0.97371} & \vi{1.077}{1.075, 1.079} & \vi{1.036}{1.034, 1.039}\\
Fixed control & 0.99095 & 0.90000 & \vi{0.90891}{0.88743, 0.92666} & \vi{0.94064}{0.90797, 0.94692} & \vi{1.153}{1.149, 1.157} & \vi{1.071}{1.067, 1.076}\\\bottomrule
\end{tabular}
\end{table}

The $(2,2)$ and $(3,4)$ filters pass the 10\% acceptance test and the duration check, with 4.29 and 4.36 fitted e-foldings, and the $(4,8)$ filter passes both on the drift-scan seeds, with individual fitted durations of 3.93-7.17 e-foldings; $(1,1)$ and $(2,4)$ fail it under both estimators. Only $(3,4)$ and $(4,8)$ also satisfy the 5\% model operating constraint. Every grid/time prediction lies inside the corresponding interval of mean individual rates. The ensemble-amplitude ratios lie above the individual-rate ratios, because the amplitude estimator weights the fastest-growing seeds (Eq.~\eqref{eq:weighting}).

For the unconstrained optimum the protected-mode signal shows no exponential growth. Over the fit window the CIC ensemble amplitude grows by a factor of 739 and the $(4,8)$ amplitude by 625, whereas the $(0,8)$ ensemble amplitude changes by a factor of 1.23, never exceeds its initial value and ends the run at 0.27 of it. Log-linear fits give a mean individual slope of 0.0025 (seed range 0.0022-0.0029)$\,\omega_p$ with $R^2<0.0013$ in every seed, 0.10-0.13 fitted e-foldings and a fit-start signal-to-floor ratio of 5.04; the duration and $R^2$ checks fail, so no rate is defined and no ratio is reported. The perturbation weights remain at most $1.5\times10^{-6}$ of the equilibrium weights. This agrees with the model status: the filter transfer $W^2=0.13143$ lies far below the marginal coupling 0.95319.

The $W^2$ column explains the ordering of the model ratios. For weak attenuation the predicted ratio is set almost entirely by the round-trip transfer at the protected mode: calibrating $1-\Rg\simeq c\,(1-W^2)$ on the selected Gaussian gives $c\simeq10.7$ and reproduces the $(2,2)$, $(3,4)$ and $(4,8)$ model values to within $3\times10^{-4}$. For the stronger filters $(1,1)$ and $(2,4)$ the response is no longer linear in $1-W^2$.

The error columns show why compensated filters give much larger error reductions. The $(3,4)$ and $(4,8)$ filters attenuate the protected mode about as much as the selected Gaussian ($W^2=0.99700$ and 0.99585 against 0.99547), yet, like every binomial filter, their transfers vanish at $k\Delta=\pi$ in both directions. The selected Gaussian has $\sigma_x=0$ and $\sigma_y=0.17\Delta_y$: it leaves the longitudinal direction unfiltered and reduces the transverse Nyquist amplitude by only 13.5\%. The difference therefore lies in the shape of the transfer function rather than in its strength at the protected mode.

For the raw four-pass binomial at $m=12$ on $N=128$, the leading low-wavenumber Gaussian approximation shown in Fig.~4(a) of the main text differs from the exact transfer by 0.512\% one-way and 1.022\% round-trip (relative to the Gaussian). All response calculations use the exact transfer.

\subsection{Weak-Landau comparison}
\label{sec:landau-design}

The Landau calculations test whether a lower density error also implies a more accurate field and an unchanged damping rate. A Maxwellian with $\vth=1$ is perturbed in mode $(1,0)$ with $k\lambda_D=0.4$ and amplitude $\alpha=0.01$, on a $64\times16$ mesh over $(5\pi,8)$, and evolved in full $f$ with velocity Verlet at $\Delta t=0.05$ and PPC 1024. Three arms share 40 paired seeds: CIC, the one-cell Gaussian $(1,1)$ and the wide Gaussian $(7.2443,7.2443)$. The wide setting is the width chosen by cross-validation of the initial density and is kept as a fixed comparator; it is not re-optimised for this experiment. Density and pre-return field errors are measured over $t\in[5,30]$ against a paired PPC-4096 CIC reference (Table~\ref{tab:landau-errors}). The coherent decay rate is fitted on the common window $[5,28]$ (Section~\ref{sec:landau-coherence}).

\begin{table}[!htbp]
\centering\small
\caption{Finite-reference error-reduction factors in the weak-Landau experiment. Values above unity indicate smaller squared error than CIC. These are errors relative to a finite PPC-4096 reference, not continuum errors (Section~\ref{sec:finite-reference}).}
\label{tab:landau-errors}
\begin{tabular}{@{}lrr@{}}\toprule
Filter & Density factor $\mathcal N_\rho$ & Field factor $\mathcal N_E$\\\midrule
One-cell Gaussian & 3.259 [3.252, 3.265] & 1.613 [1.603, 1.622]\\
Wide Gaussian & 4.612 [4.598, 4.625] & 0.632 [0.619, 0.646]\\\bottomrule
\end{tabular}
\end{table}

A field factor below unity means a larger field error than CIC: the wide filter increases it by $1/0.632\simeq1.58$. The CIC baseline covers only 1.58 fitted e-foldings on the common window, below the two-e-folding requirement of Section~\ref{sec:checks}. The one-cell coherent-rate interval lies inside the 10\% tolerance, but neither filtered arm can be fully confirmed against such a short baseline. The coherent rate also contains a contribution from loss of phase coherence between seeds, which Section~\ref{sec:landau-coherence} separates from the decay of individual amplitudes.

\section{Particle loading and initial perturbation}
\label{sec:loading}

This section specifies the initial states. The projected-linear two-stream calculations (drift scan, exact-width test and binomial rates) are designed so that seeds differ only in their velocity samples. Velocities come from a fixed quantile template, so every seed contains the same set of thermal velocity values and differs only in how these values are assigned to markers. Marker positions and initial weights are the same for every seed and every drift, so the initial amplitude and phase of the target mode are identical across seeds. This is the loading used in this study, not a general requirement of PIC.

\paragraph{Positions}
With $N_{x,p}=2048$ and $N_{y,p}=512$ lattice points, positions are the cell centres of a regular particle lattice,
\begin{equation}
 x_{ab}=(a+\tfrac12)L_x/N_{x,p},\qquad
 y_{ab}=(b+\tfrac12)L_y/N_{y,p},\qquad a=0,\dots,N_{x,p}-1,\ b=0,\dots,N_{y,p}-1.
 \label{eq:positions}
\end{equation}
This gives $N_p=1{,}048{,}576$ markers, or 512 per mesh cell (16 along $x$ and 32 along $y$).

\paragraph{Velocities}
Define the quantile template $z_j=\Phi^{-1}(\xi_j)$ with $\xi_j=(j+\tfrac12)/N_p$, $j=0,\dots,N_p-1$, where $\Phi$ is the standard normal cumulative distribution function. Two random permutations $\pi_x$ and $\pi_y$ assign
\begin{equation}
 v^{\rm th}_{x,j}=\vth z_{\pi_x(j)},\qquad v_{y,j}=\vth z_{\pi_y(j)}.
\end{equation}
A third permutation assigns exactly $N_p/2$ beam signs $\beta_j=+1$ and $N_p/2$ signs $\beta_j=-1$, and
\begin{equation}
 v_{x,j}=v^{\rm th}_{x,j}+\beta_j\vb .
 \label{eq:loading}
\end{equation}
The three permutations are successive draws from one random stream, so a seed fixes their joint realisation, including the sample correlation between beam assignment and thermal velocity. These correlations are what distinguish one seed from another, and the finite-marker model of Section~\ref{sec:finite-marker} retains them.

\paragraph{Weights and perturbation}
Each marker carries the state $(\xx_j,\vv_j,w^{\rm full}_j,w^\delta_j)$, where $w^{\rm full}_j=-L_xL_y/N_p$ is its equilibrium charge and $w^\delta_j$ its signed perturbation charge. For amplitude $\alpha=10^{-8}$ the initial perturbation is
\begin{equation}
 w^{\delta}_j(0)=-\alpha w^{\rm full}_j\cos(\kk\cdot\xx_j),
 \qquad
 w^{\delta}_j(0)\leftarrow w^{\delta}_j(0)-\frac{1}{N_p}\sum_l w^{\delta}_l(0),
 \label{eq:initial-weight}
\end{equation}
where the second step removes the residual zero mode of the finite marker set. All arms at a given drift and seed start from the same positions, velocities, beam signs and weights; only the filter changes.

\paragraph{Full-$f$ loading}
The two-stream reconstruction calculations use the same lattice construction with $4\times8$ markers per cell at PPC 32 and $32\times32$ at PPC 1024, and the velocity loading of Eq.~\eqref{eq:loading}. The $(12,0)$ perturbation is imposed by displacing the markers, $x\to x+\alpha k_x\sin(k_xx)/k_x^2$ with $\alpha=10^{-3}$. Each PPC-1024 reference is generated separately, with its own marker count and permutations, so it is not a refinement of the PPC-32 sample. The weak-Landau calculations load a single Maxwellian from one global quantile template on a $32\times32$-per-cell lattice (PPC 1024) and impose the $(1,0)$ perturbation through a position displacement of amplitude $\alpha=0.01$ together with the matching first-order velocity shift of the Langmuir mode, so that density and current are both perturbed. The projected Landau calculation of Section~\ref{sec:landau-df} uses the same positions and velocities with fixed equilibrium weights and perturbation weights
\begin{equation}
 w^\delta_j(0)=w^{\rm full}_j\,\alpha\Big[-\cos\varphi_j+\frac{v_{x,j}}{k\vth^2}\big(\gamma_L\sin\varphi_j-\omega_L\cos\varphi_j\big)\Big],\qquad \varphi_j=kx_j-\varphi_0,
 \label{eq:landau-weights}
\end{equation}
followed by removal of the mean, where $\omega_L-\ii\gamma_L=1.285057-0.066128\ii$ is the continuum root. The second term is the first-order velocity shift of the forward Langmuir wave. Each seed is run with $\varphi_0=0$ and $\varphi_0=\pi/2$, and the two target-mode coefficients are combined as $\widehat E_{\cos}+\ii\widehat E_{\sin}$. Because the evolution is linear, this combination is the response to the complex initial perturbation $\ee^{\ii kx}$, for which the position-velocity pairing of the markers cancels from the principal free-streaming term (Section~\ref{sec:landau-df}).

\section{Particle-field coupling and \texorpdfstring{$\delta f$}{delta-f} update}
\label{sec:coupling}

This section writes out the operators that the response models of Section~\ref{sec:response} describe. Together they implement the matched coupling of Eq.~(5) of the main text for the projected $\delta f$ calculation.

\paragraph{Deposition, filtering and field solve}
Let $C_{gj}(\xx_j)$ be the periodic bilinear CIC coefficient from marker $j$ to node $g$, with $\sum_gC_{gj}=1$, and let $\Delta x=L_x/N_x$, $\Delta y=L_y/N_y$. Deposition and gathering use the same coefficients~\cite{birdsall2004plasma,hockney1988computer}. Only the perturbation charge is deposited, since the ion background cancels the equilibrium charge:
\begin{equation}
 \rho_g=\frac{1}{\Delta x\Delta y}\sum_j C_{gj}w^{\delta}_j.
 \label{eq:deposit}
\end{equation}
For a mesh quantity $u_g$ we use $\widehat u_\kk=(N_xN_y)^{-1}\sum_g u_g\,\ee^{-\ii\kk\cdot\xx_g}$. A real, even, separable filter with transfer $W_\kk$ acts on the deposited density,
\begin{equation}
 \widehat\rho^{\,d}_\kk=W_\kk\,\widehat\rho_\kk .
 \label{eq:deposit-filter}
\end{equation}
All Fourier coefficients are then set to zero except those of the target $+\kk$ and its conjugate $-\kk$. Keeping both returns a real density; removing all other modes ensures that the measured growth belongs to the target mode alone and that no competing instability can develop. The field follows from
\begin{equation}
 \widehat\phi_\kk=\frac{\widehat\rho^{\,d}_\kk}{|\kk|^2},\qquad
 \widehat\EE_\kk=-\ii\kk\,\widehat\phi_\kk,\qquad \kk\ne0,
 \label{eq:poisson}
\end{equation}
with the zero mode set to zero. The return path applies the same filter and then CIC interpolation,
\begin{equation}
 \EE_j=\sum_gC_{gj}\left\{\mathcal F^{-1}\big[W_\kk\widehat\EE_\kk\big]\right\}_g .
 \label{eq:gather}
\end{equation}
A resolved mode therefore carries one factor $W_\kk$ in the field before the return filter and $W_\kk^2$ in the complete particle-force coupling.

\paragraph{Transfer functions}
The CIC shape is the convolution of two top-hats in each direction, so its transfer is~\cite{hockney1988computer,langdon1970effects}
\begin{equation}
 S_\CIC(\qq)=\operatorname{sinc}^2\!\left(\frac{q_x\Delta x}{2}\right)
 \operatorname{sinc}^2\!\left(\frac{q_y\Delta y}{2}\right),
 \qquad \operatorname{sinc}z=\frac{\sin z}{z},
 \label{eq:cic-shape}
\end{equation}
and matched CIC contributes $S_\CIC^2$ to the force response. Here $\qq$ is a general wavevector; in Section~\ref{sec:gridtime} it also denotes a spatial alias. The Gaussian filter has
\begin{equation}
 W_H(\kk)=\exp\!\big[-\tfrac12\{(k_x\sigma_x)^2+(k_y\sigma_y)^2\}\big].
 \label{eq:gaussian}
\end{equation}
The one-pass binomial stencil $(1,2,1)/4$ has transfer $\cos^2(k\Delta/2)=1-s$ with $s=\sin^2(k\Delta/2)$ along the filtered coordinate~\cite{birdsall2004plasma,hockney1988computer}. Applying it $n$ times and multiplying by the degree-$p$ Taylor polynomial of $(1-s)^{-n}$ gives the compensated filter
\begin{equation}
 W_{p,n}=(1-s)^n\sum_{r=0}^{p}\binom{n+r-1}{r}s^r=1+\mathcal O(s^{p+1}),
 \label{eq:binomial}
\end{equation}
with $W_{0,n}=(1-s)^n$ the raw filter. The commonly quoted width $\sigma_{\rm eq}=\sqrt{n/2}\,\Delta$ is only the leading small-$k\Delta$ match to the raw binomial (Appendix~A of the main text).

\paragraph{Linearised weight equation}
Writing $f=f_0+\delta f$ and keeping terms linear in the perturbation, markers move along unperturbed straight-line orbits and carry the perturbation in their weights. Because markers sample $f_0$, dividing the source term by $f_0$ produces the velocity score
\begin{equation}
 g_x(\vv)=\partial_{v_x}\log f_0=-\frac{v_x-\vb\tanh(v_x\vb/\vth^2)}{\vth^2},\qquad
 g_y(\vv)=\partial_{v_y}\log f_0=-\frac{v_y}{\vth^2}
 \label{eq:score}
\end{equation}
for the two-beam mixture of Eq.~\eqref{eq:f0}, which equals a constant times $\exp[-(v_x^2+v_y^2+\vb^2)/(2\vth^2)]\cosh(v_x\vb/\vth^2)$. Multiplying by the equilibrium charge gives the projected linearised update~\cite{parker1993fully,hu1994generalized,denton1995bounded}
\begin{equation}
 \dot w^{\delta}_j=-\frac{q_e}{m_e}(\EE_j\cdot\gvec_j)\,w^{\rm full}_j,
 \qquad \dot\xx_j=\vv_j,\qquad \dot\vv_j=0 .
 \label{eq:weight-ode}
\end{equation}
Since velocities are fixed, this calculation contains no velocity-space heating.

\paragraph{Time integration}
The update uses the explicit midpoint method. With $w_j^n=w_j^\delta(t_n)$ and $\R_j(\xx,w)$ the right-hand side of Eq.~\eqref{eq:weight-ode},
\begin{equation}
\begin{aligned}
 \xx_j^{n+1/2}&=\xx_j^n+\tfrac12\Delta t\,\vv_j, &
 w_j^{n+1/2}&=w_j^n+\tfrac12\Delta t\,\R_j(\xx^n,w^n),\\
 \xx_j^{n+1}&=\xx_j^n+\Delta t\,\vv_j, &
 w_j^{n+1}&=w_j^n+\Delta t\,\R_j(\xx^{n+1/2},w^{n+1/2}).
\end{aligned}
 \label{eq:midpoint}
\end{equation}
Positions are wrapped periodically. The field is recomputed from the half-step positions and weights before the second stage, and the mean of $w^\delta$ is removed after each stage. The full-$f$ Landau calculations instead use velocity Verlet; Section~\ref{sec:continuum} shows that this changes the reference damping rate negligibly.

\section{Response models}
\label{sec:response}

Three models of increasing specificity predict the filtered rate (Table~\ref{tab:hierarchy}). All use the exact filter transfer. They differ in how they treat the mesh, the time step and the particle sample, and therefore answer different questions. The continuous-time principal-harmonic model explains the mechanism. The continuum-velocity grid/time model is the one used to set constraints. The finite-marker model explains why individual seeds grow at different rates.

\begin{table}[!htbp]
\centering\small
\caption{Response models used in the main text and in this supplement. Equation and figure numbers without the prefix S refer to the main text.}
\label{tab:hierarchy}
\begin{tabularx}{\linewidth}{@{}P{0.2\linewidth}P{0.16\linewidth}P{0.15\linewidth}P{0.15\linewidth}L@{}}\toprule
Model & Space & Time & Velocities & Used for\\\midrule
Continuous-time principal harmonic (Section~\ref{sec:continuum}; Eq.~(8)) & Principal harmonic with $S_\CIC^2W^2$ & Continuous & Continuum $f_0$ & Mechanism; Fig.~1; full-$f$ Landau and longitudinal two-stream model rates; Penrose check\\\addlinespace
Continuum-velocity grid/time (Section~\ref{sec:gridtime}) & Spatial aliases, $|a_i|\le8$ & Explicit midpoint, temporal aliases $|l|\le1024$ & Continuum $f_0$ & Interior widths, 10\% boundary, selection constraint, root status, binomial and projected-Landau predictions\\\addlinespace
Seed-specific finite marker (Section~\ref{sec:finite-marker}) & Principal harmonic & Explicit midpoint & The $N_p$ loaded velocities and beam signs of one seed & Seed-to-seed spread; Fig.~\ref{fig:marker}; Section~\ref{sec:drift-results}\\\bottomrule
\end{tabularx}
\end{table}

\subsection{Continuous-time principal-harmonic model}
\label{sec:continuum}

Let $Z(\zeta)=\pi^{-1/2}\int_{-\infty}^{\infty}\ee^{-u^2}/(u-\zeta)\dd u$ for $\operatorname{Im}\zeta>0$, continued analytically elsewhere~\cite{fried1961dispersion}, and let $\kappa=|\kk|$. For the mixture of Eq.~\eqref{eq:f0} define
\begin{equation}
 \zeta_\pm=\frac{\omega\mp k_x\vb}{\sqrt2\,\kappa\vth},\qquad
 \X(\omega,\kk)=\tfrac12\big[1+\zeta_+Z(\zeta_+)\big]+\tfrac12\big[1+\zeta_-Z(\zeta_-)\big].
\end{equation}
Integrating the linearised Vlasov equation over velocity gives the susceptibility $\X/(\kappa^2\vth^2)$~\cite{fried1961dispersion,adams2023petscpic}, so the unsmoothed continuum root solves
\begin{equation}
 \epsilon_{\rm eq}(\omega,\kk)=1+\frac{\X(\omega,\kk)}{\kappa^2\vth^2}=0.
 \label{eq:continuum}
\end{equation}
For a single Maxwellian, $\X$ is replaced by $1+\zeta Z(\zeta)$ with $\zeta=\omega/(\sqrt2\kappa\vth)$. This root is the rate of the ideal homogeneous equilibrium; it contains neither the particle shape nor the filter. Multiplying the susceptibility by the principal round-trip factor $S_\CIC^2W^2$ gives the continuous-time principal-harmonic relation, Eq.~(8) of the main text, which underlies Fig.~1 of the main text and the Landau model ratios.

\paragraph{Time discretisation in the Landau calculation}
The full-$f$ Landau runs use velocity Verlet. Linearising its position recurrence $x^{n+1}-2x^n+x^{n-1}=(q_e/m_e)\Delta t^2E^n$ about ballistic orbits replaces the continuous-time susceptibility by a sum over temporal aliases~\cite{hockney1988computer},
\begin{equation}
 D_{\rm V}(k,\omega)=1+W_k^2S_\CIC^2(k)\sum_{l\in\mathbb Z}\chi_L\!\left(k,\omega-\frac{2\pi l}{\Delta t}\right),\qquad
 \chi_L=\frac{1+\zeta Z(\zeta)}{k^2\vth^2}.
 \label{eq:verlet-alias}
\end{equation}
At $k=0.4$, $\vth=1$, $\Delta x=5\pi/64$ and $\Delta t=0.05$, the principal-CIC damping rate is $-0.066285\,\omega_p$ for velocity Verlet and $-0.066308\,\omega_p$ in continuous time, a relative difference below 0.04\%. Since every Landau measurement is a ratio to a CIC run with the same integrator, the continuous-time model is used for the predicted Landau ratios.

\subsection{Continuum-velocity, grid/time-discrete model}
\label{sec:gridtime}

A finite mesh and time step modify the principal-harmonic relation in two ways. Sampling a continuous shape on the mesh repeats its Fourier response at reciprocal-grid wavevectors~\cite{langdon1970effects,huang2016finite,meyers2015dispersion}, and a discrete update repeats the frequency response at multiples of $2\pi/\Delta t$. The model of this section keeps the continuum velocity distribution but includes both effects for the explicit-midpoint update of Section~\ref{sec:coupling}.

The spatial aliases of a represented mode are
\begin{equation}
 \qq_{\mathbf a}=\kk+2\pi\left(\frac{a_x}{\Delta x},\frac{a_y}{\Delta y}\right),\qquad \mathbf a=(a_x,a_y)\in\mathbb Z^2 .
\end{equation}
With time dependence $\exp(\lambda t)$, $\lambda=-\ii\omega$, the temporal aliases are $\omega_l=\ii\lambda-2\pi l/\Delta t$, $l\in\mathbb Z$, and the velocity-integrated response of alias $l$ is
\begin{equation}
 \J_l(\lambda)=-\sum_{\mathbf a}
 \frac{W_\kk^2\,S_\CIC^2(\qq_{\mathbf a})\,(\kk\cdot\qq_{\mathbf a})}
 {|\kk|^2|\qq_{\mathbf a}|^2\vth^2}\,
 \X(\omega_l,\qq_{\mathbf a}),
 \label{eq:J}
\end{equation}
where the two-stream shifts use $q_{a,x}\vb$ and each $\zeta_\pm$ uses $|\qq_{\mathbf a}|$. With $\tau=\Delta t/2$, the half-step source contributes
\begin{equation}
 c_0=-\sum_{\mathbf a}W_\kk^2S_\CIC^2(\qq_{\mathbf a})
 \frac{\kk\cdot\qq_{\mathbf a}}{|\kk|^2}\,\tau\,
 \exp\!\big[-\tfrac12\vth^2|\qq_{\mathbf a}|^2\tau^2\big]\cos(q_{a,x}\vb\tau),
\end{equation}
and the growth rate is the root of
\begin{equation}
 \R_{\rm grid,time}(\lambda)=1-\sum_l\J_l(\lambda)
 -\frac{\Delta t}{2}\,c_0\,\ee^{-\lambda\Delta t/2}\sum_l(-1)^l\J_l(\lambda)=0.
 \label{eq:gridtime}
\end{equation}

\paragraph{Derivation}
Equation~\eqref{eq:gridtime} follows from the midpoint map of Section~\ref{sec:finite-marker}, with the principal harmonic generalised to all spatial aliases and the marker sums replaced by equilibrium velocity integrals. For the complex streaming variable $x=\lambda+\ii\qq_{\mathbf a}\cdot\vv$, the partial-fraction identities~\cite{dlmfhyperbolic}
\begin{equation}
 \frac{\Delta t}{\ee^{x\Delta t}-1}=\sum_{l=-\infty}^{\infty}\frac{1}{x+2\pi\ii l/\Delta t}-\frac{\Delta t}{2},\qquad
 \frac{\Delta t}{2\sinh(x\Delta t/2)}=\sum_{l=-\infty}^{\infty}\frac{(-1)^l}{x+2\pi\ii l/\Delta t},
 \label{eq:temporal-identities}
\end{equation}
taken with symmetric summation, convert the discrete-time denominators into sums over temporal aliases. The velocity integral of the unphased score vanishes, so the constant term in the first identity drops out, and the first marker sum becomes $\sum_l\J_l$. The half-step sum becomes $\ee^{-\lambda\Delta t/2}\sum_l(-1)^l\J_l$, and integrating its score by parts gives $c_0$, including the Gaussian streaming factor and the symmetric-beam cosine. The result is the response of this midpoint update, not a generic leapfrog dielectric.

\paragraph{Relation to the other models}
The continuum root of Eq.~\eqref{eq:continuum} contains neither the CIC shape nor discrete-time phases. The continuous-time principal-harmonic model adds only the principal spatial shape. Equation~\eqref{eq:gridtime} adds spatial aliases and the midpoint time response. Setting the temporal-alias index to zero inside the midpoint expression truncates the aliases; it does not by itself recover the continuous-time limit. Because the velocity sum has been replaced by an integral, Eq.~\eqref{eq:gridtime} describes the discretised operators acting on an ideal equilibrium rather than any particular finite marker set. Faithful treatment of aliases, update phases and the operators actually used is known to matter in PIC dispersion analysis~\cite{meyers2015dispersion,huang2016finite}.

\paragraph{Convergence}
The calculations truncate the alias sums at $|a_x|,|a_y|\le8$ and $|l|\le1024$. Table~\ref{tab:alias-convergence} shows that this truncation is converged far below the measurement uncertainty: increasing the cutoffs to $(10,2048)$ changes the CIC rate by $1.08\times10^{-8}\,\omega_p$ and the selected-width ratio by $8.7\times10^{-9}$. These numbers quantify truncation only; they are not estimates of finite-marker or other model error.

\begin{table}[!htbp]
\centering\small
\caption{Alias-cutoff convergence of the grid/time response at $\vb=0.80$, $(m_x,m_y)=(12,1)$ and the selected Gaussian $(h_x,h_y)=(0,0.1716393)$. Spatial indices obey $|a_x|,|a_y|\le a_{\max}$ and temporal indices $|l|\le l_{\max}$.}
\label{tab:alias-convergence}
\begin{tabular}{@{}rrrr@{}}\toprule
$a_{\max}$ & $l_{\max}$ & $\gamma_{\CIC}/\omega_p$ & $\gamma_*/\gamma_{\CIC}$\\\midrule
0 & 1024 & 0.11744402510 & 0.95127980293\\
2 & 1024 & 0.11746407576 & 0.95130096642\\
4 & 1024 & 0.11746394862 & 0.95130085989\\
8 & 512 & 0.11746391475 & 0.95130083233\\
8 & 1024 & 0.11746393707 & 0.95130085030\\
10 & 2048 & 0.11746394786 & 0.95130085898\\\bottomrule
\end{tabular}
\end{table}

\subsection{Seed-specific finite-marker model}
\label{sec:finite-marker}

The grid/time model predicts one rate for the ideal equilibrium, but different seeds grow at measurably different rates (Section~4.4 of the main text). The finite-marker model asks whether this spread is already determined by the initial velocity sample. It replaces the continuum velocity integral by a sum over the $N_p$ markers actually loaded, keeping the principal $+\kk$ harmonic and the midpoint phases. In effect it is the dispersion relation of a plasma composed of exactly the sampled velocities.

For each seed, the velocities and beam signs $\{v_{x,j},v_{y,j},\beta_j\}$ are generated by the loading of Section~\ref{sec:loading}, which is deterministic once the seed is fixed. The model is therefore evaluated on the same initial marker set as the simulation of that seed; the initial mode amplitude and phase obtained from this marker set agree with those of the simulations. Define
\begin{equation}
 \Omega_j=\kk\cdot\vv_j,\qquad
 u_j=\ee^{-\ii\Omega_j\Delta t},\qquad \psi_j=\ee^{-\ii\Omega_j\Delta t/2},\qquad
 a_j=W_\kk^2\,\frac{\ii S_\CIC^2(\kk)}{|\kk|^2N_p}\,(\kk\cdot\gvec_j),\qquad
 c_N=\sum_j\psi_ja_j .
\end{equation}
Here $u_j$ and $\psi_j$ are the full- and half-step streaming phases of marker $j$, $a_j$ is its feedback per unit summed modal amplitude, and $c_N$ is the half-step phase-weighted sum of these feedbacks.

Let $b_j^n$ be the demodulated perturbation amplitude of marker $j$, including the factor $\ee^{-\ii\kk\cdot\xx_j^n}$, so that free streaming multiplies it by $u_j$, and let $B^n=\sum_jb_j^n$. The two midpoint stages of Eq.~\eqref{eq:midpoint}, restricted to the principal mode, read
\begin{equation}
\begin{aligned}
 b_j^{n+1/2}&=\psi_jb_j^n+\tfrac{\Delta t}{2}\psi_ja_jB^n,\qquad
 B^{n+1/2}=\sum_\ell \psi_\ell b_\ell^n+\tfrac{\Delta t}{2}c_NB^n,\\
 b_j^{n+1}&=u_jb_j^n+\Delta t\,\psi_ja_jB^{n+1/2},
\end{aligned}
 \label{eq:modal-midpoint}
\end{equation}
where the factor $\psi_j$ in the last line carries the midpoint source to the full-step phase. Eliminating the half step gives the one-step map
\begin{equation}
 b_j^{n+1}=u_jb_j^n+\Delta t\,\psi_ja_j\sum_\ell\Big(\psi_\ell+\tfrac{\Delta t}{2}c_N\Big)b_\ell^n.
 \label{eq:modal-map}
\end{equation}
For a collective mode $b_j^n=B_jz^n$ with $z=\ee^{\lambda\Delta t}$, define $F=\sum_\ell(\psi_\ell+\Delta t\,c_N/2)B_\ell$. Then $(z-u_j)B_j=\Delta t\,\psi_ja_jF$. Substituting this into the definition of $F$, dividing by $F\ne0$ and using $\psi_j^2=u_j$ gives
\begin{equation}
 \R_N(\lambda)=1-\Delta t\sum_j\frac{u_ja_j}{\ee^{\lambda\Delta t}-u_j}
 -\frac{\Delta t^2}{2}\,c_N\sum_j\frac{\psi_ja_j}{\ee^{\lambda\Delta t}-u_j}=0.
 \label{eq:finite-marker}
\end{equation}
Replacing the marker sums by velocity integrals recovers the principal-harmonic part of Eq.~\eqref{eq:gridtime}. All seeds contain the same set of velocity values (Section~\ref{sec:loading}), but they distribute these values differently between the two beams and between $v_x$ and $v_y$; the sets $\{\Omega_j,a_j\}$, and hence the roots of Eq.~\eqref{eq:finite-marker}, therefore differ from seed to seed.

\paragraph{Root finding}
The root of Eq.~\eqref{eq:finite-marker} is found by complex Newton iteration started from the grid/time root and converged to a residual below $2\times10^{-11}$. At $\vb=0.818$ the fixed control's transfer lies below the grid/time marginal coupling (Section~\ref{sec:root-status}); there the finite-marker root is continued from the unfiltered root, again without reference to any measured rate.

\paragraph{Approximations}
Equation~\eqref{eq:finite-marker} is the principal-mode, diagonal approximation of the finite-marker midpoint map, not the exact eigenvalue of the complete discrete system. It omits spatial aliases, the finite-sample coupling between $+\kk$ and $-\kk$, and any nonlinear or full-$f$ response. Section~\ref{sec:drift-results} shows that, despite these omissions, it reproduces the measured seed-to-seed variation closely.

\subsection{Model inputs and independence from measured rates}
\label{sec:inputs}

A prediction can test a measurement only if it does not use it. The three models take as inputs only the mesh and domain $(N_x,N_y,L_x,L_y)$, the mode indices $(m_x,m_y)$, the time step, the equilibrium parameters $(\vb,\vth)$, the exact filter transfer $W_\kk$ and the alias cutoffs. The finite-marker model additionally uses the loaded velocities and beam signs of each seed. Measured mode amplitudes, fit windows and fitted rates never enter Eqs.~\eqref{eq:continuum}, \eqref{eq:gridtime} or \eqref{eq:finite-marker}; every prediction follows from the operators and the initial state alone.

\subsection{Root status and marginal coupling}
\label{sec:root-status}

The constraint compares a filtered root with the CIC root of the same branch, so each candidate is classified before a ratio is formed. The branch is continued from the CIC root in steps of $10^{-3}$ in the round-trip transfer $a=W^2$, each step starting Newton's method from the previous root. A candidate is \emph{growing} if the continuation reaches its transfer with $\operatorname{Re}\lambda>0$; its ratio is then $\Rg=\operatorname{Re}\lambda/\gamma_\CIC$. It is \emph{marginal} if its transfer lies below the coupling $a_c$ at which the tracked growth rate reaches zero. It is \emph{unresolved} if the continuation fails before either outcome. A failed Newton iteration is never read as zero growth, and only growing candidates can be admissible.

For the symmetric two-beam equilibrium the unstable branch of the protected mode is purely growing, $\lambda$ real, and the residual is real for real $\lambda$ (its imaginary part is below $10^{-31}$). The marginal coupling is the transfer at which the tracked growth rate reaches zero,
\begin{equation}
 \R_{\rm grid,time}(\lambda\to0^+;a_c)=0 .
 \label{eq:marginal}
\end{equation}
At $\vb=0.80$ the grid/time model gives $a_c=0.953187$, and the continued branch confirms it: $\gamma=0.0380$, 0.0291 and 0.0156\,$\omega_p$ at $a=0.958$, 0.956 and 0.954. With only the principal harmonic and continuous time, Eq.~\eqref{eq:marginal} gives $a_c=0.953156$.

The marginal point is not an exact double root. Near $(\lambda,a)=(0,a_c)$ the residual expands as
\begin{equation}
 \R(\lambda,a)\simeq\R_a(a-a_c)+\R_\lambda\lambda+\tfrac12\R_{\lambda\lambda}\lambda^2 ,
 \label{eq:marginal-expansion}
\end{equation}
and the grid/time model gives $\R_a=-1.049$, $\R_{\lambda\lambda}=6.969$ and $\R_\lambda=9.26\times10^{-5}\,\omega_p^{-1}$, small but nonzero. In the continuous-time principal-harmonic model $\R_\lambda$ is proportional to the second derivative of the projected distribution at the origin, which is exponentially small, of order $\exp[-u_b^2/(2\vth^2)]\sim10^{-50}$, for these widely separated beams; the spatial and temporal aliases of the grid/time model contribute the larger value. Because $\R_\lambda\neq0$, the two real roots collide at a coupling $a_c-\R_\lambda^2/(2|\R_a|\R_{\lambda\lambda})=a_c-5.9\times10^{-10}$ and at $\lambda=-\R_\lambda/\R_{\lambda\lambda}=-1.3\times10^{-5}\,\omega_p$, not at $\lambda=0$. The pure square-root law $\gamma=[2|\R_a|(a-a_c)/\R_{\lambda\lambda}]^{1/2}$, the reactive limit underlying Eq.~(10) of the main text, therefore holds where the quadratic term dominates the linear one, $\lambda\gg2\R_\lambda/\R_{\lambda\lambda}=2.7\times10^{-5}\,\omega_p$ or $|a-a_c|\gg2\times10^{-9}$. At $a=0.954$ it differs from the root of Eq.~\eqref{eq:marginal-expansion} by $1.3\times10^{-5}\,\omega_p$, or 0.085\%, so over the whole range relevant to the selection the branch follows the square-root behaviour closely without being exactly described by it arbitrarily close to marginality. None of this affects $a_c$, which is defined by Eq.~\eqref{eq:marginal} directly, or the classification below.

The principal-harmonic result can be checked independently. Its dispersion relation depends on the velocity distribution only through the projection $F(u)$ of $f_0$ onto $\hat\kk=\kk/|\kk|$. For the present equilibrium $F$ is the sum of two Gaussians centred at $\pm u_b$, $u_b=\vb k_x/|\kk|=0.75895$, with standard deviation $\vth$, and has a single minimum at $u=0$. The Penrose criterion~\cite{penrose1960electrostatic} then states that a growing root exists if and only if
\begin{equation}
 a\,S_\CIC^2(\kk)\int_{-\infty}^{\infty}\frac{F(u)-F(0)}{u^2}\,\dd u>|\kk|^2 .
 \label{eq:penrose}
\end{equation}
With $\int[F(u)-F(0)]u^{-2}\dd u=1.75922$ and $S_\CIC^2=0.91968$, the threshold is $a_c=0.953155$, in agreement with the principal-harmonic root of Eq.~\eqref{eq:marginal}. For $a<a_c$ the principal-harmonic model therefore has no growing root at all, not merely no root on the tracked branch. The grid/time model, which adds spatial aliases and the midpoint response, moves the threshold by $3\times10^{-5}$.

Of the 30 binomial candidates, 17 are growing and 13 have transfers below $a_c$: all six raw filters, $(1,n)$ for $n\ge2$, and $(2,6)$ and $(2,8)$. The closest to the threshold is $(1,2)$ with $W^2=0.95208$. No candidate is unresolved. All 84 Gaussians have $W^2\ge0.99095$ and are growing. At the terminal drift $\vb=0.818$, $a_c=0.99715$, above the fixed control's transfer 0.99095, so the fixed control is marginal there.

\section{Reconstruction-error objective and constrained selection}
\label{sec:selection}

This section specifies the statistical half of the selection rule, Eqs.~(12)-(13) of the main text: the candidate set, the error objective and the resulting choice. The response constraint that defines the admissible set comes from the grid/time model of Section~\ref{sec:gridtime}.

\subsection{Candidate design}
\label{sec:candidates}

The Gaussian design at $\vb=0.80$ consists of CIC and 84 Gaussians. The Gaussians occupy four attenuation levels at the protected mode, $q/q_{\max}=0.25$, 0.50, 0.75 and 1, where $q_{\max}=0.00909$ is the 10\% boundary of the grid/time model (Eq.~(14) of the main text). Each level contains 21 allocations of the attenuation between the two directions, with allocation fraction $k_x^2\sigma_x^2/q$ running from 0 (purely transverse smoothing) to 1 (purely longitudinal). Because the model response depends on the filter only through $W_\kk^2$ at the protected mode, all candidates on one level share a single predicted rate ratio, 0.97595, 0.95130, 0.92600 and 0.90000 respectively. The binomial family adds the 30 compensated filters of Section~\ref{sec:binomial-design}; their model ratios follow from their exact $W^2$ in the same way, and 13 of them have transfers below the marginal coupling $a_c=0.95319$, so their tracked branch has reached marginality (Section~\ref{sec:root-status}). The 5\% operating constraint, $\Rg\ge0.95$, admits CIC, the 42 Gaussians on the two weakest levels and 11 binomial filters.

\subsection{Snapshot error objective}
\label{sec:objective}

The objective is a paired-reference reconstruction error. Full-$f$ simulations at PPC 32 are evolved with CIC coupling and $\Delta t=0.05$ from an initial $(12,0)$ density perturbation of amplitude $10^{-3}$. At times $t\in[0.5,4]$, spaced by 0.1, each candidate filter is applied to the CIC deposit of these particle states, and the result is compared with the CIC deposit of a paired, independently generated PPC-1024 reference simulation. The density error is Eq.~(13) of the main text; the field error uses the squared Euclidean difference of the nodal field before the return filter. All candidates therefore see identical particle data, and CIC is simply the zero-width member of each family. The objective measures how well each filter reconstructs the density from a given particle sample; it does not compare trajectories evolved with different filters. Five seeds select the filter and five disjoint seeds evaluate it.

This objective differs from least-squares cross-validation, which estimates the integrated squared error of a kernel estimate by leaving out each particle in turn~\cite{silverman1986density,wand1995kernel,qin2018kde}. The usual unbiasedness argument for cross-validation assumes independent samples and does not carry over automatically to quiet-start or otherwise correlated PIC loads. A directly measured reference error avoids this question at the cost of a reference simulation.

\paragraph{Where the error lies}
For the loading used here and a representative seed, the time-averaged CIC error power is spread over all resolved wavelengths: 38\% lies in the lower half of both wavenumber ranges, 43\% at $|k_x|$ above half the Nyquist wavenumber, and only 0.42\% in the protected mode. If the candidate-reference covariance is small and the sampling variance scales inversely with particle count, the reference contributes about $1/(1+32)\simeq3\%$ of the CIC pair error (Section~\ref{sec:finite-reference}), which caps any reduction factor near 33. The selected filters are far from this ceiling, while the raw eight-pass factor, 21.7, is about two thirds of it. The inverse-count scaling is an assumption for these lattice loads, so the ceiling is approximate.

\subsection{Selection, evaluation and stability}
\label{sec:selection-result}

\begin{table}[!htbp]
\centering\small
\caption{Constrained selection at $\vb=0.80$. Error factors are $\mathcal N=\widehat{\mathcal E}(\CIC)/\widehat{\mathcal E}(\theta)$ on the five evaluation seeds, with 95\% intervals from 5,000 bootstrap resamples of those seeds. }
\label{tab:selection}
\setlength{\tabcolsep}{4pt}
\begin{tabular}{@{}lrrrr@{}}\toprule
Selection & Filter & Model $\Rg$ & Density $\mathcal N_\rho$ & Field $\mathcal N_E$\\\midrule
Gaussian, 5\% & $(0,0.1716)$ & 0.95130 & 1.0772 [1.0752, 1.0791] & 1.0363 [1.0341, 1.0386]\\
Gaussian, 10\% & $(0,0.2427)$ & 0.90000 & 1.1527 [1.1486, 1.1566] & 1.0710 [1.0666, 1.0757]\\
Binomial, 5\% & $(p,n)=(4,8)$ & 0.95548 & 4.2442 [4.0081, 4.4886] & 1.6026 [1.5441, 1.6703]\\
All, 5\% & $(p,n)=(4,8)$ & 0.95548 & 4.2442 [4.0081, 4.4886] & 1.6026 [1.5441, 1.6703]\\
All, unconstrained & $(p,n)=(0,8)$ & marginal & 21.7273 [20.5225, 22.9988] & 6.7893 [6.3182, 7.3558]\\\bottomrule
\end{tabular}
\end{table}

Table~\ref{tab:selection} lists the selected filters. Within the admissible Gaussians, the density error decreases monotonically with attenuation and is smallest when all of it is transverse. The density and field objectives both select $(h_x,h_y)=(0,0.1716393)$, on the strongest admissible level. This choice leaves the excited longitudinal $(12,0)$ signal unattenuated while smoothing transverse fluctuations; it reflects this snapshot ensemble and is not a general rule. The excluded 10\%-boundary Gaussian $(0,0.242735)$ would give a density factor 1.1527~[1.1486, 1.1566], or 13.24\% lower MSE.

Over the full candidate set, both objectives select the $(4,8)$ compensated filter. Its density error on the evaluation seeds is 76.44\% below CIC, compared with 7.16\% for the selected Gaussian. Without the constraint, both objectives select the raw eight-pass filter $(0,8)$, whose tracked branch is marginal (Section~\ref{sec:root-status}); its density error is 95.40\% below CIC. These are reconstruction errors of offline-filtered CIC snapshots against the finite reference. They contain variance, bias and reference contributions (Section~\ref{sec:finite-reference}) and are not noise-free continuum errors.

A nested bootstrap that resamples both the selection and the evaluation seeds selects the same candidate in all 5,000 draws for each family, with density intervals [1.0752, 1.0791] (Gaussian) and [4.0081, 4.5103] ($(4,8)$). With five selection seeds the bootstrap can produce at most $\binom{9}{5}=126$ distinct selection samples, and an exact enumeration of all 126 selects the same filters. The selection is therefore stable within the available seeds, although five seeds cannot establish stability against new realisations.

\section{Rate estimators, uncertainty and acceptance checks}
\label{sec:estimators}

\subsection{Estimators}
For seed $s$ and a fixed window $I$, the four rates used in the paper are
\begin{equation}
\begin{aligned}
 \lambda_{\rm eq}&:\ \epsilon_{\rm eq}(\ii\lambda_{\rm eq},\kk)=0, &
 \lambda_{N,s}&:\ \R_{N,s}(\lambda_{N,s})=0,\\
 \widehat\gamma_s&=\operatorname{slope}_{t\in I}\log|\widehat E_s(t)|, &
 \widehat\gamma_{\rm amp}&=\operatorname{slope}_{t\in I}\log\Big[N_s^{-1}\sum_s|\widehat E_s(t)|\Big],
\end{aligned}
 \label{eq:estimands}
\end{equation}
where $N_s$ is the number of seeds, $\widehat E_s$ is the target-mode Fourier coefficient of $E_x$, and each slope is an ordinary least-squares fit on $I$. They answer different questions: $\lambda_{\rm eq}$ characterises the ideal homogeneous equilibrium; $\lambda_{N,s}$ predicts the principal-mode response of one initial marker sample; $\widehat\gamma_s$ measures the growth of one simulation; and $\widehat\gamma_{\rm amp}$ measures the growth of the seed-averaged amplitude. The mean individual rate is $\overline\gamma=N_s^{-1}\sum_s\widehat\gamma_s$.

The two measured quantities differ systematically. For positive amplitudes $|E_s|=A_s\exp(\gamma_st)$,
\begin{equation}
 \frac{\dd}{\dd t}\log\sum_sA_s\ee^{\gamma_st}
 =\frac{\sum_s\gamma_sA_s\ee^{\gamma_st}}{\sum_sA_s\ee^{\gamma_st}},
 \label{eq:weighting}
\end{equation}
so faster-growing seeds receive increasing weight and $\widehat\gamma_{\rm amp}$ generally exceeds $\overline\gamma$ for a growing mode. For a damped mode the same weighting favours the slowest decay. Neither estimator is claimed to be unbiased for the equilibrium rate; they are reported separately and compared with the model that predicts each.

\subsection{Uncertainty}
Ratios are formed from paired arms: a filtered run and its CIC baseline share a seed. Intervals come from the paired nonparametric bootstrap~\cite{efron1979bootstrap}: seeds are resampled 5,000 times with replacement, keeping each candidate with its CIC baseline, the complete ratio is recomputed in each draw, and the 2.5th and 97.5th percentiles are reported.

\subsection{Acceptance checks}
\label{sec:checks}
A filter preserves the rate if the paired 95\% interval of its rate ratio lies wholly within $[0.90,1.10]$. This statistical test is applied together with three checks that ensure the rate is measurable.
\begin{enumerate}
\item \emph{Duration.} The fit must span at least two amplitude e-foldings,
\begin{equation}
 N_e=|\widehat\gamma|(t_2-t_1)\ge2,
 \label{eq:duration-check}
\end{equation}
for both the baseline and the candidate, with rate and window referring to the same observable.
\item \emph{Signal.} For the two-stream calculations, the ensemble amplitude at the start of the fit must exceed its noise floor by at least a factor of five, and the ensemble fit must satisfy $R^2\ge0.98$. The floor is the smallest median of the ensemble amplitude over complete two-time-unit windows before the fit start; it measures the level of the initial transient that precedes exponential growth.
\item \emph{Linearity.} The signed perturbation weights must remain small compared with the equilibrium weights,
\begin{equation}
 \frac{\max_{j,t}|w_j^{\delta}(t)|}{\langle|w_j^{\rm full}|\rangle}<0.5,
 \label{eq:weight-check}
\end{equation}
over all markers and the full simulation. This is an operational small-perturbation criterion, not a general $\delta f$ error bound.
\end{enumerate}
Section~\ref{sec:exact-width} adds individual-fit quality checks for the exact-width test.

\subsection{Fit-window sensitivity}
\label{sec:window}
The fit windows were fixed before the measurements, and the primary window is kept throughout the paper. Because the seed intervals contain only seed-sampling uncertainty, the dependence of the measured ratios on the window is reported separately here. Table~\ref{tab:window} refits the arms at $\vb=0.80$ that use the drift-scan seeds on three alternative windows: the fit start delayed by 10 and by 20 time units with the same endpoint, and the endpoint advanced by 10. All refits use the same seeds and pairing; the shifted-window intervals use separate bootstrap draws.

\begin{table}[!htbp]
\centering\small
\caption{Ratios of mean individual rates at $\vb=0.80$ on the primary window $[24.32,69.16]$ and three alternative windows, with paired 95\% seed-bootstrap intervals. ``Model'' is the grid/time prediction. The last row gives the mean CIC fitted e-foldings on each window.}
\label{tab:window}
\setlength{\tabcolsep}{3pt}
\begin{tabular}{@{}lrrrrr@{}}\toprule
Setting & Model & $[24.32,69.16]$ & $[34.32,69.16]$ & $[44.32,69.16]$ & $[24.32,59.16]$\\\midrule
Fixed control & 0.90000 & \vi{0.90891}{0.88743, 0.92666} & \vi{0.90663}{0.88315, 0.92550} & \vi{0.90614}{0.88187, 0.92531} & \vi{0.91019}{0.88853, 0.92742}\\
Selected Gaussian & 0.95130 & \vi{0.95565}{0.94527, 0.96401} & \vi{0.95462}{0.94383, 0.96345} & \vi{0.95439}{0.94331, 0.96340} & \vi{0.95622}{0.94643, 0.96434}\\
$(4,8)$, selected & 0.95548 & \vi{0.95945}{0.94995, 0.96702} & \vi{0.95853}{0.94806, 0.96634} & \vi{0.95831}{0.94755, 0.96631} & \vi{0.95997}{0.95064, 0.96722}\\
\addlinespace
CIC e-foldings & -- & 5.58 & 4.32 & 3.08 & 4.34\\\bottomrule
\end{tabular}
\end{table}

Delaying the fit start lowers every ratio, by 0.10-0.13\% for the selected filters and by 0.25-0.30\% for the fixed control; advancing the endpoint raises them by 0.05-0.14\%. These shifts are smaller than the seed intervals, which are wide because of the seed-to-seed spread of Section~\ref{sec:drift-results}, but they are systematic: all ratios move in the same direction, so the seed intervals do not represent this uncertainty. For the selected Gaussian and $(4,8)$ filters every window gives a ratio well inside the 10\% acceptance band. The fixed control, which lies on the acceptance limit by design, has an interval crossing 0.90 in every window, so its classification remains unresolved.

\section{Drift-scan measurements and finite-marker predictions}
\label{sec:drift-results}

Table~\ref{tab:estimands} lists both measured ratios and the finite-marker prediction for every arm of the drift scan. Three features stand out. First, the fixed control's effect grows steadily as the drift approaches marginal stability, from a 0.7\% reduction at $\vb=0.60$ to 9.1\% at $\vb=0.80$, whereas the interior setting stays close to its 0.95 target at every drift. Second, the finite-marker prediction tracks the mean individual-rate ratio to within $2\times10^{-3}$ in every resolved row. Third, the ensemble-amplitude ratio departs from the mean individual-rate ratio by an amount that grows with drift; at $\vb=0.80$ the difference is large enough to change the acceptance decision for the fixed control.

\begin{table}[!htbp]
\centering\small
\caption{Filtered/CIC rate ratios in the drift scan. Brackets are paired 95\% seed-bootstrap intervals. The finite-marker column is computed from the initial state alone (Section~\ref{sec:finite-marker}). The $\vb=0.818$ rows fail the signal and linearity checks of Section~\ref{sec:checks} and are descriptive only.}
\label{tab:estimands}
\begin{tabular}{@{}llrrr@{}}\toprule
$\vb$ & Setting & Ensemble amplitude & Mean individual rates & Finite marker\\\midrule
0.60 & Fixed & 0.99278 [0.99268, 0.99287] & 0.99275 [0.99267, 0.99284] & 0.99275\\
     & Interior & 0.95059 [0.94991, 0.95122] & 0.95038 [0.94979, 0.95103] & 0.95035\\
0.70 & Fixed & 0.98595 [0.98537, 0.98636] & 0.98567 [0.98525, 0.98611] & 0.98570\\
     & Interior & 0.95173 [0.94970, 0.95317] & 0.95074 [0.94927, 0.95228] & 0.95085\\
0.76 & Fixed & 0.97269 [0.96971, 0.97426] & 0.97057 [0.96860, 0.97254] & 0.97048\\
     & Interior & 0.95546 [0.95052, 0.95804] & 0.95193 [0.94868, 0.95518] & 0.95179\\
0.80 & Fixed & 0.94064 [0.90797, 0.94692] & 0.90891 [0.88743, 0.92666] & 0.90702\\
     & Interior & 0.96986 [0.95376, 0.97301] & 0.95446 [0.94407, 0.96310] & 0.95360\\\addlinespace
0.818 & Fixed & 0.89249 & 0.57955 & 0.56219\\
      & Interior & 0.99689 & 0.98751 & 0.98797\\\bottomrule
\end{tabular}
\end{table}

\begin{figure}[!htbp]
\centering\includegraphics[width=\linewidth]{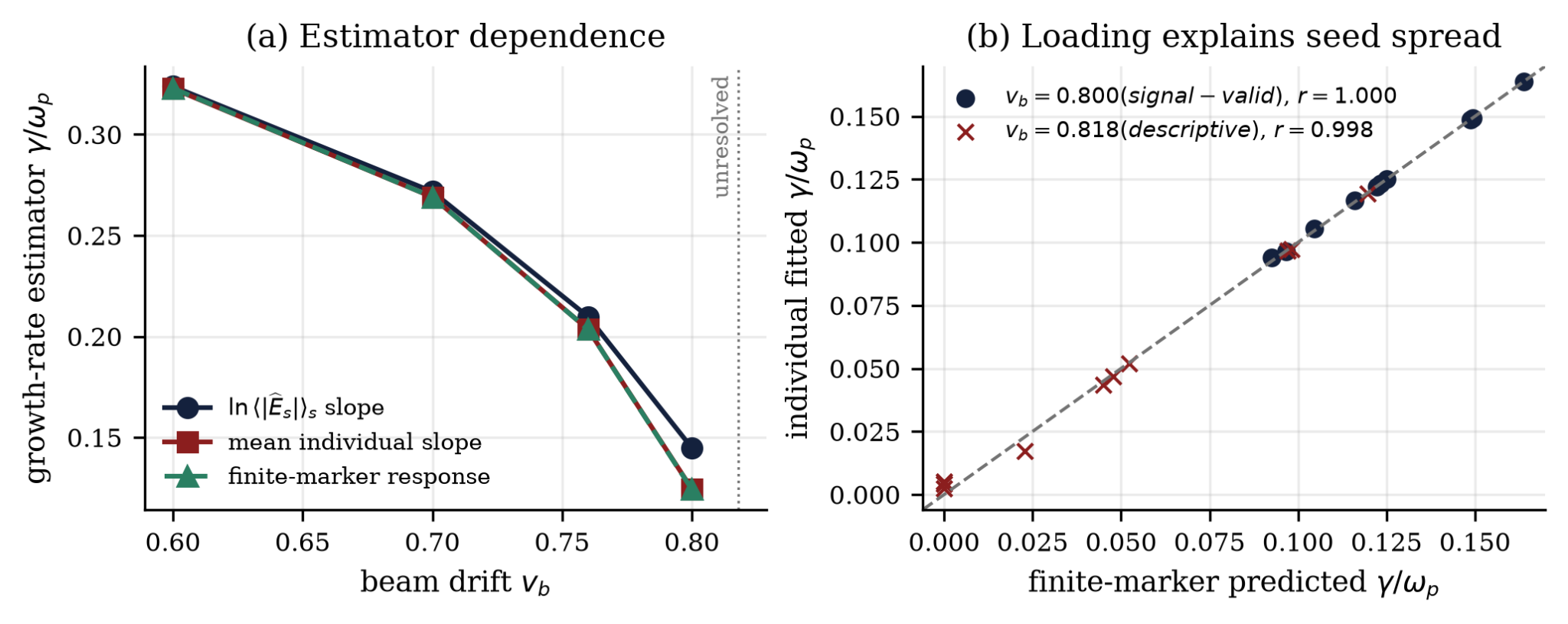}
\caption{Finite-marker predictions and estimator dependence for the CIC baseline. (a) Growth-rate estimators across the resolved drifts: slope of the logarithm of the seed-averaged amplitude, mean of the individual seed slopes, and mean of the finite-marker predictions. The amplitude slope exceeds the mean individual rate increasingly as the drift grows, while the finite-marker mean follows the individual rates. (b) Individual fitted rates against finite-marker predictions at $\vb=0.80$ (circles; $r=0.999885$) and at the unresolved drift $\vb=0.818$ (crosses), which fails the signal and linearity checks and is shown for completeness. The dashed line is equality.}
\label{fig:marker}
\end{figure}

\paragraph{Acceptance at $\vb=0.80$}
The fixed-control individual-rate interval $[0.88743,0.92666]$ crosses the lower limit 0.90. These data therefore do not establish preservation within 10\%, nor do they establish a larger distortion. Its ensemble-amplitude interval $[0.90797,0.94692]$ lies inside the tolerance band, because the amplitude estimator weights the fastest-growing seeds most heavily (Eq.~\eqref{eq:weighting}). The interior interval lies inside the band for both estimators.

\paragraph{Seed-by-seed agreement}
For the 120 simulations that pass all checks ($\vb=0.60$, 0.70, 0.76 and 0.80; three arms of ten seeds each), the correlation between the finite-marker prediction and the individual fitted rate lies between 0.9982 and 0.99994. At $\vb=0.80$ the mean observed-minus-predicted residuals are $2.53\times10^{-4}$ (CIC), $4.64\times10^{-4}$ (fixed) and $3.49\times10^{-4}$ (interior), with seedwise root-mean-square errors of $4.88\times10^{-4}$, $8.91\times10^{-4}$ and $6.56\times10^{-4}$, all in units of $\omega_p$. These residuals are small compared with the seed-to-seed spread of the CIC rates (standard deviation $0.0233\,\omega_p$ about a mean of $0.1245\,\omega_p$). The spread is therefore a property of the initial velocity sample, captured by Eq.~\eqref{eq:finite-marker}, rather than of fitting noise or of the time integration (Fig.~\ref{fig:marker}b). It characterises the sampled $\delta f$ calculation at this marker number, not the growth rate of the physical equilibrium; finite-sample variability of this kind is expected in PIC ensembles~\cite{touati2022ensemble}. The residuals measure the error of the approximations listed in Section~\ref{sec:finite-marker} and are not folded into the acceptance intervals.

\section{Landau envelopes, phase coherence and finite-reference error}
\label{sec:landau}

\subsection{Coherent envelope and phase coherence}
\label{sec:landau-coherence}

A damped mode can be followed only until it reaches the noise floor. To lower the floor, the Landau analysis averages the complex mode coefficient over seeds before fitting. This coherent average decays for two reasons: the individual amplitudes damp, and their phases drift apart so that they cancel more strongly when added. This section describes how the fit window is set and then separates the two contributions.

\paragraph{Envelope and fit window}
The coherent magnitude and its geometric moving-average envelope are
\begin{equation}
 A_n=\Big|N_s^{-1}\sum_s\widehat E_s(t_n)\Big|,\qquad
 \widetilde A_n=\exp\!\left[\frac1{49}\sum_{r=-24}^{24}\ln A_{n+r}\right],
 \label{eq:envelope}
\end{equation}
where $n$ indexes output times spaced by 0.1, so that the 49-sample window spans about one oscillation period (4.8894). The envelope removes the oscillation at $\omega_r$ while keeping the slow decay. The floor of each arm is the smallest five-unit rolling median of the coherent signal within $[70,90]$, a late interval in which the damped signal has reached the noise level. A fit begins at $t=5$ and ends before the envelope persistently falls below five times the floor: the crossing is the first run of five consecutive samples below that level, and the fit ends before the crossing minus the 24-sample half-width of the envelope. This rule gives end times of 30.1 (CIC), 35.6 (one-cell) and 28 (wide). All three arms are fitted on the common window $[5,28]$, the shortest of the three, so that every arm is measured over identical times.

\paragraph{Phase coherence}
With $\widehat E_s$ the complex mode coefficient of seed $s$, the amplitude-weighted phase coherence is
\begin{equation}
 C(t)=\frac{|\sum_s\widehat E_s(t)|}{\sum_s|\widehat E_s(t)|}.
 \label{eq:coherence}
\end{equation}
It equals one when all seeds share the same phase and decreases as their phases separate. The coherent magnitude is exactly $C(t)$ times the mean individual magnitude.

\paragraph{An illustrative model of rate and frequency spread}
Consider seeds $E_s(t)=A\exp[(\mu_\gamma+\delta\gamma_s)t-\ii(\omega_0+\delta\omega_s)t]$ with common real amplitude and phase and independent zero-mean Gaussian offsets of variance $\sigma_\gamma^2$ and $\sigma_\omega^2$. The Gaussian moment-generating and characteristic functions give
\begin{equation}
 M_\infty(t)=\mathbb E|E_s(t)|=A\,\ee^{\mu_\gamma t+\sigma_\gamma^2t^2/2},\qquad
 C_\infty(t)=\frac{|\mathbb E E_s(t)|}{\mathbb E|E_s(t)|}=\ee^{-\sigma_\omega^2t^2/2}.
\end{equation}
For a uniformly weighted log-linear fit on $[t_1,t_2]$, $\operatorname{Cov}(t,t^2)/\operatorname{Var}(t)=t_1+t_2$, so
\begin{equation}
 \gamma_{M,\infty}=\mu_\gamma+\tfrac12\sigma_\gamma^2(t_1+t_2),\qquad
 \gamma_{C,\infty}=-\tfrac12\sigma_\omega^2(t_1+t_2),\qquad
 \gamma_{{\rm coh},\infty}=\mu_\gamma+\tfrac12(\sigma_\gamma^2-\sigma_\omega^2)(t_1+t_2).
 \label{eq:gaussian-spreads}
\end{equation}
A spread in rates biases the mean-magnitude slope towards slower decay, a spread in frequencies adds a negative coherence slope, and equal spreads cancel in the coherent slope. The model is illustrative: it is not fitted to the present data and omits correlations, transients and finite-sample effects. The decomposition below requires none of its assumptions.

\paragraph{Three-arm decomposition on a common window}
For each 40-seed ensemble define $M(t)=N_s^{-1}\sum_s|\widehat E_s(t)|$ and $A(t)=|N_s^{-1}\sum_s\widehat E_s(t)|$, so that $A=MC$. Let $\mathcal G$ be the 49-sample moving average in log space and $\mathcal L_I$ the least-squares slope on $I=[5,28]$. Both operators are linear, so
\begin{equation}
 \mathcal L_I\mathcal G\log A=\mathcal L_I\mathcal G\log M+\mathcal L_I\mathcal G\log C,
 \qquad\text{i.e.}\qquad \widehat\gamma_{\rm coh}=\widehat\gamma_M+\widehat\gamma_C .
 \label{eq:phase-accounting}
\end{equation}
This is an algebraic identity for the measured observable, satisfied numerically to $1.4\times10^{-16}\,\omega_p$, and is independent of the Gaussian model above. Table~\ref{tab:landau-windows} gives the three terms for each arm.

\begin{table}[!htbp]
\centering\small
\caption{Three-arm Landau fits on the common window $[5,28]$. Rates are in units of $\omega_p$; ratio intervals are paired 95\% seed-bootstrap intervals. The CIC and one-cell fits span fewer than two e-foldings.}
\label{tab:landau-windows}
\begin{tabular}{@{}lrrr@{}}\toprule
Quantity & CIC & One-cell Gaussian & Wide Gaussian\\\midrule
Coherent rate $\widehat\gamma_{\rm coh}$ & $-0.068791$ & $-0.070039$ & $-0.129881$\\
Mean-magnitude term $\widehat\gamma_M$ & $-0.065958$ & $-0.066994$ & $-0.088064$\\
Coherence term $\widehat\gamma_C$ & $-0.002833$ & $-0.003045$ & $-0.041816$\\
Coherent-decay ratio to CIC & 1 & 1.01814 & 1.88805\\
95\% interval & -- & [1.01486, 1.02121] & [1.66753, 2.14374]\\
Mean-magnitude ratio to CIC & 1 & 1.01570 & 1.33515\\
Principal-harmonic prediction & 1 & 1.01635 & 1.97986\\
Fitted e-foldings & 1.582 & 1.611 & 2.987\\
Envelope coherence at $t=28$ & 0.9247 & 0.9183 & 0.3523\\\bottomrule
\end{tabular}
\end{table}

For CIC and the one-cell filter the coherence term is small, and the coherent rate is essentially the mean-magnitude rate. For the wide filter it is not: its phase coherence falls to 0.35 by $t=28$ (Fig.~5(d) of the main text), and the coherence term carries the fraction
\begin{equation}
 F_C=\frac{\widehat\gamma_{C,\rm wide}-\widehat\gamma_{C,\CIC}}
 {\widehat\gamma_{{\rm coh},\rm wide}-\widehat\gamma_{{\rm coh},\CIC}}
 =0.6381\;[0.5745,0.6878]
\end{equation}
of the wide-minus-CIC difference in coherent slope. The interval uses 5,000 paired resamples across all three arms, recomputing the complete reduction in each draw. $F_C$ describes how the measured slope difference is composed; it is not a causal attribution or a fraction of an eigenvalue shift.

Neither term alone estimates the damping eigenvalue. The mean-magnitude slope is amplitude-weighted towards the slowest-decaying seeds (Eq.~\eqref{eq:weighting}), while dephasing adds a negative coherence slope. The principal-harmonic predictions, 1.01635 and 1.97986, lie inside the one-cell and wide coherent-ratio intervals, whereas the mean-magnitude ratios are 1.01570 and 1.33515. The agreement of the coherent rate with the model is therefore compatible with the decomposition: as Eq.~\eqref{eq:gaussian-spreads} illustrates, opposing shifts in the two terms can leave the coherent slope close to a model eigenvalue. The one-cell filter meets the coherent-rate criterion on this window and the wide filter does not; the short CIC baseline still prevents full confirmation.

\subsection{Finite-reference error decomposition}
\label{sec:finite-reference}

All error reductions in the paper are measured against a reference simulation with more particles, not against the exact density. Let $X_H$ and $R$ be the candidate and reference densities at one node and time, regarded as random variables across seeds, with means $\overline X_H$ and $\overline R$. Using population-normalised variances and covariances,
\begin{equation}
 \operatorname{MSE}(X_H-R)=(\overline X_H-\overline R)^2+\operatorname{Var}(X_H)+\operatorname{Var}(R)-2\operatorname{Cov}(X_H,R).
 \label{eq:mse-decomposition}
\end{equation}
The first term is the squared bias relative to the reference mean, the next two are the sampling variances of candidate and reference, and the last is their covariance. All terms are averaged over the same nodes and times, and the same identity applies componentwise to the field before the return filter. The reference variance is shared by every candidate, so it sets a floor that no filter can remove.

\paragraph{Reference-limited ceiling}
In the Landau comparison, reference variance accounts for 19.50\%, 63.54\% and 89.93\% of the density pair MSE for CIC, the one-cell filter and the wide filter. If the candidate-reference covariance is negligible, no filter can then reduce the CIC density error by more than a factor of about $1/0.195\simeq5.1$. The wide filter's factor of 4.612 already reaches about 90\% of this ceiling, so its density advantage over the one-cell filter is compressed by the reference noise. The 19.5\% share is close to the value $0.25/(1+0.25)=0.20$ expected if sampling variance scales inversely with particle count and the reference has four times as many particles. The same scaling gives about 3\% for the two-stream reconstruction calculations, where the reference has 32 times as many particles; the selected two-stream reductions are then far from their reference-limited ceiling (Section~\ref{sec:objective}).

\paragraph{Variance-adjusted error}
Removing the reference variance and covariance defines
\begin{equation}
 \mathcal E^{\rm adj}_H=\operatorname{MSE}(X_H-R)-\operatorname{Var}(R)+2\operatorname{Cov}(X_H,R)
 =\operatorname{Var}(X_H)+(\overline X_H-\overline R)^2,
 \label{eq:Q}
\end{equation}
the squared error of the candidate relative to the reference mean. For the Landau density, $\mathcal E^{\rm adj}_\CIC/\mathcal E^{\rm adj}_H$ is 7.218 for the one-cell Gaussian and 36.858 for the wide Gaussian; the corresponding field ratios are 1.8951 and 0.5803. The adjustment strengthens the density ordering but leaves the field ordering unchanged: the wide filter still has the larger field error. These adjusted ratios are descriptive. No intervals are given for them, because the paired bootstrap of a variance-subtracted quantity is biased: its expected shift is $[\widehat{\operatorname{Var}}(R)-2\widehat{\operatorname{Cov}}(X_H,R)]/N_s$ node by node, which does not yield a simple correction for a ratio. The main text therefore reports the directly measured pair error.

\section{Dynamics test of the selected filters and the unconstrained optimum}
\label{sec:exact-width}

This section gives the protocol and quality checks for the dynamics tests of Section~4.2 of the main text. The selection itself is described in Section~\ref{sec:selection}.

\paragraph{Protocol}
The selected Gaussian $(h_x,h_y)=(0,0.1716393)$, the selected $(4,8)$ compensated filter and the raw eight-pass filter $(0,8)$ are each applied throughout the projected-linear evolution at $\vb=0.80$ (mode $(12,1)$, PPC 512, $\Delta t=0.02$, final time 71.16) for the ten drift-scan seeds. These seeds are disjoint from both the selection and evaluation samples of Section~\ref{sec:selection}. Each filtered run starts from the same initial state as the drift-scan CIC run of the same seed (Section~\ref{sec:loading}), which serves as its paired baseline. The calculation is deterministic for a given seed: an independent recomputation of a test seed reproduces the CIC and selected-Gaussian histories to a relative difference below $10^{-10}$. The seed set, the fit window $[24.32,69.16]$ and the primary observable, the ratio of mean individual rates with its paired interval required to lie in $[0.90,1.10]$, are those of the drift scan, and no seed is added or removed. The ensemble-amplitude ratio is the secondary observable.

\paragraph{Quality checks}
All checks pass for the selected filters. The arm-level fit-start signal-to-floor ratios are 22.2 (CIC), 22.4 (selected Gaussian) and 22.4 ($(4,8)$), far above the required five. Every individual fit also satisfies $R^2\ge0.98$ (minimum 0.9919 for CIC, 0.9863 for the selected Gaussian and 0.9869 for $(4,8)$), every individual rate is positive, and individual fitted durations span 4.21-7.35 e-foldings for CIC, 3.90-7.16 for the selected Gaussian and 3.93-7.17 for $(4,8)$. The largest perturbation-weight ratios are 0.016, 0.012 and 0.012, well below the bound of 0.5. Some filtered seeds have individual signal-to-floor ratios slightly below five; they are retained because the signal criterion applies to the arm-level ensemble amplitude.

\paragraph{Result}
For the selected Gaussian the primary ratio is 0.955645 [0.945273, 0.964006], inside the acceptance band and consistent with the grid/time prediction 0.95130. For $(4,8)$ it is 0.959452 [0.949954, 0.967020], consistent with the prediction 0.95548. The secondary ensemble-amplitude ratios, 0.970634 [0.955356, 0.973707] and 0.973122 [0.959121, 0.975947], are higher because the amplitude estimator weights the fastest-growing seeds (Eq.~\eqref{eq:weighting}). Both filters pass the 10\% acceptance test under both estimators.

\paragraph{Unconstrained optimum}
The raw eight-pass filter $(0,8)$ was run with the same protocol, seeds and window. Its protected-mode signal shows no resolved growth (Section~\ref{sec:binomial-design}): the ensemble amplitude changes by a factor of 1.23 across the window against 739 for CIC, the fits span 0.10-0.13 e-foldings with $R^2<0.0013$, and the duration and fit-quality checks fail, so no rate ratio is defined. The measurement agrees with the model status of Section~\ref{sec:root-status}. It shows that the unconstrained density optimum does not preserve the protected linear mode.

\section{Projected-linear Landau damping}
\label{sec:landau-df}

This calculation measures the damping rates of the one-cell and wide Gaussians against a CIC baseline that spans more than two e-foldings, which the full-$f$ comparison of Section~\ref{sec:landau} cannot provide.

\paragraph{Configuration}
The mesh, domain, marker lattice and velocities are those of the full-$f$ Landau runs: $64\times16$ cells on $(5\pi,8)$, PPC 1024 ($N_p=2^{20}$ markers), $\vth=1$, $k\lambda_D=0.4$. Markers move at fixed velocities, and the perturbation weights follow Eq.~\eqref{eq:weight-ode} with the Maxwellian score $\gvec=-\vv/\vth^2$, advanced by the explicit midpoint method of Eq.~\eqref{eq:midpoint} with $\Delta t=0.05$ to $t=70$. Only the $(\pm1,0)$ Fourier modes are retained before the field solve. The filter is active in both directions of the coupling; its transfer at the target mode is $W=0.99519$ (one-cell) and $W=0.77654$ (wide), so $W^2=0.99041$ and 0.60301. Three seeds are used, each paired across the three arms. The initial weights are Eq.~\eqref{eq:landau-weights}.

\paragraph{Why a travelling-wave perturbation}
In the projected calculation the target-mode coefficient is a sum over markers. In the principal, diagonal approximation, which keeps the response of the $+k$ harmonic to itself and omits the coupling between $+k$ and $-k$ through the finite marker sample, the part driven by the field depends on each marker only through $\exp[-\ii k v_{x,j}(t-t')]$, because the phase $\ee^{\ii kx_j}$ acquired at the source time cancels against the deposit phase at the observation time. In that approximation it is set by the velocity template alone and is the same for every seed. A finite sample that retains both harmonics also couples them, through a kernel of the form
\begin{equation}
 K_{+-}(t,t')\propto\sum_j(\kk\cdot\gvec_j)\,\ee^{-\ii\kk\cdot[\xx_j(t)+\xx_j(t')]},
\end{equation}
which vanishes under uniform spatial averaging but not for a finite sample with randomly paired positions and velocities. This is the same finite-sample $\pm k$ coupling that the finite-marker model of Section~\ref{sec:finite-marker} omits, and it is present before spatial aliases are added. The free-streaming part of a $\cos kx$ perturbation is different. Writing $\cos\varphi_j=\tfrac12(\ee^{\ii\varphi_j}+\ee^{-\ii\varphi_j})$, it contains
\begin{equation}
 \tfrac12\sum_jh_j\,\ee^{-\ii kv_{x,j}t}+\tfrac12\sum_jh_j\,\ee^{-2\ii kx_j}\ee^{-\ii kv_{x,j}t},
 \qquad h_j=w^{\rm full}_j\Big[1+\frac{\omega_L v_{x,j}}{k\vth^2}+\dots\Big],
\end{equation}
up to the sign conventions of Eq.~\eqref{eq:landau-weights}. The first sum is again set by the template. The second vanishes on average but, because positions and velocities are paired by a random permutation, it is a random sum of order $(\sum_j|h_j|^2)^{1/2}$ that does not decay. Relative to the initial amplitude this residual is a few times $10^{-3}$ for $N_p=2^{20}$, enhanced by the velocity term of the eigenmode weights; the field response to it adds a further contribution. With a $\cos kx$ perturbation alone, the ten-seed mean amplitude of the wide arm levels off at about $7\times10^{-3}$ of its initial value after $t\approx40$, and the individual CIC rates on $[5,35]$ vary by 4\% between seeds. The $\sin kx$ perturbation produces the same template sum multiplied by $-\ii$ and the second sum with the opposite sign, so the combination $\widehat E_{\cos}+\ii\widehat E_{\sin}$ cancels this specified initial free-streaming term exactly in the principal mode. It does not remove the finite-sample $\pm k$ coupling of the field-driven response, nor the contributions of spatial aliases, which carry the factor $S_\CIC(\qq_{\mathbf a})$. In the wide arm the amplitude at $t=70$ is 1.4-$4.8\times10^{-4}$ of its initial value, of which the decayed eigenmode accounts for about $1\times10^{-4}$; we have not separated these two sources, which would require a direct Fourier deposit without CIC aliases, and attribute the residual to their combination.

\paragraph{Estimators}
The initial weights excite mainly the forward Langmuir wave, but also a weaker backward wave with the same damping, whose amplitude is 0.18 (CIC and one-cell) and 0.31 (wide) of the forward one. Their beat modulates $|\widehat E|$ at $2\omega_L$ (Fig.~6(a) of the main text), so log-amplitude slopes depend slightly on where the window ends relative to the beat. The primary estimator is a two-exponential fit to the complex coefficient on $t\in[8,65]$ by the matrix-pencil method~\cite{hua1990matrix}, which returns both roots; the forward root has the larger amplitude. Its relative residual over the whole window is at most $1.6\times10^{-3}$ in every run, but this figure is dominated by the larger early signal. Over $t\ge40$ alone the relative residual is 1.2\% (CIC and one-cell) and 9.5\% (wide), where the wide signal approaches the residual level. Ordinary least-squares log-amplitude slopes are reported on $[5,35]$ and $[5,50]$. On $[5,65]$ the wide arm reaches the residual, which biases its slope towards slower decay (ratio 1.91), so this window is not used for the slope ratios; the two-exponential fit, which weights the larger early signal, responds differently to the same contamination, and its sensitivity is tested directly below.

\paragraph{Sensitivity of the root fit}
We repeated the two-exponential fit with start times 5, 8 and 12, end times 45, 55, 65 and 70, and $M=2$, 3 and 4 exponentials (36 settings). The forward CIC rate lies between $-0.066431$ and $-0.066391\,\omega_p$ (prediction $-0.066362$). The one-cell ratio lies between 1.01628 and 1.01641 and the wide ratio between 1.97406 and 1.97837 (predictions 1.01633 and 1.97883). Across these choices the CIC rate varies by 0.06\%, the one-cell ratio by 0.01\% and the wide ratio by 0.22\%, even though the late-window residual of the wide arm grows to 13\% when the fit extends to $t=70$.

\begin{table}[!htbp]
\centering\small
\caption{Projected-linear Landau roots and damping ratios (mean of three seeds). Roots are $\lambda=\gamma-\ii\omega$ in units of $\omega_p$; the predictions are the grid/time roots of the midpoint update (Section~\ref{sec:gridtime}). Fitted e-foldings and the smallest $R^2$ over seeds refer to the log-amplitude slope; $R^2$ is reduced by the forward-backward beat.}
\label{tab:landau-df}
\setlength{\tabcolsep}{4pt}
\begin{tabular}{@{}lrrr@{}}\toprule
Quantity & CIC & One-cell Gaussian & Wide Gaussian\\\midrule
$W^2$ at $(1,0)$ & 1 & 0.99041 & 0.60301\\
Predicted root & $-0.066362-1.28428\,\ii$ & $-0.067446-1.28061\,\ii$ & $-0.131320-1.11409\,\ii$\\
Measured forward root, $[8,65]$ & $-0.066401-1.28427\,\ii$ & $-0.067487-1.28059\,\ii$ & $-0.131293-1.11411\,\ii$\\
Damping ratio, root fit & 1 & 1.01636 & 1.97728\\
Damping ratio, slope $[5,35]$ & 1 & 1.01578 & 1.97892\\
Damping ratio, slope $[5,50]$ & 1 & 1.01673 & 1.97521\\
Predicted ratio & 1 & 1.01633 & 1.97883\\\addlinespace
E-foldings, $[5,35]$ and $[5,50]$ & 1.998, 2.99 & 2.03, 3.04 & 3.95, 5.90\\
Smallest $R^2$, $[5,35]$ and $[5,50]$ & 0.955, 0.979 & 0.955, 0.979 & 0.963, 0.983\\\bottomrule
\end{tabular}
\end{table}

\paragraph{Results}
The measured roots agree with the grid/time predictions to within 0.07\% in the damping rate and $2\times10^{-5}$ in the frequency for every arm (Table~\ref{tab:landau-df}). The root-fit damping ratios differ from the predictions by $+0.003\%$ (one-cell) and $-0.08\%$ (wide), and the slope ratios on the two windows lie within 0.2\% of them. The per-seed root-fit ratios range over 1.01619-1.01646 and 1.97527-1.97952. The CIC baseline spans 1.998 e-foldings on $[5,35]$, marginally short of two, and 2.99 on $[5,50]$ and 3.8 on the root-fit window, so the duration requirement of Section~\ref{sec:checks} is met on the latter two windows. The one-cell filter changes the damping rate by 1.6\%, inside the 10\% tolerance, and the wide filter nearly doubles it. With only three seeds that agree closely, no bootstrap interval is quoted; the spread between estimators, windows and fit settings, up to 0.25\%, is the more relevant uncertainty. The three seeds share the prescribed velocity template and differ only in its permutations and position pairing, so their agreement tests the permutation-sensitive effects; it does not establish convergence with respect to the velocity sample. These are linear-response measurements of the damping eigenvalue. The full-$f$ coherent ratios of Section~\ref{sec:landau-coherence}, 1.01814 and 1.88805, are consistent with them within their intervals and additionally contain the dephasing between seeds.

\setlength{\bibsep}{2pt}
\bibliographystyle{elsarticle-num}
\bibliography{references}